\documentclass[preprint,aps,floatfix,superscriptaddress,pre]{revtex4}
\usepackage{amsmath, amsthm, amssymb, graphicx}
\usepackage{color}
\usepackage{multirow}
\usepackage{ulem}

\input epsf.sty
\begin{document}

\title{Behavior analysis of virtual item gambling}
\author{Xiangwen Wang}
\affiliation{Department of Physics, Virginia Tech, Blacksburg, VA 24061-0435, USA}
\affiliation{Center for Soft Matter and Biological Physics, Virginia Tech, Blacksburg, VA 24061-0435, USA}
\affiliation{Department of Statistics, Virginia Tech, Blacksburg, VA 24061-0439, USA}
\affiliation{Department of Computer Science, Virginia Tech, Blacksburg, VA 24061-0106, USA}
\author{Michel Pleimling}
\affiliation{Department of Physics, Virginia Tech, Blacksburg, VA 24061-0435, USA}
\affiliation{Center for Soft Matter and Biological Physics, Virginia Tech, Blacksburg, VA 24061-0435, USA}
\affiliation{Academy of Integrated Science, Virginia Tech, Blacksburg, VA 24061-0405, USA}
\date{\today}

\begin{abstract}
From the gambling logs of an online lottery game we extract the probability distribution of various
quantities (e.g., bet value, total pool size, waiting time between successive gambles) as well as related correlation
coefficients. We view the net change of income of each player as a random walk. The mean squared
displacement of these net income random walks exhibits a transition between a super-diffusive and
a normal diffusive regime. We discuss different random walk models with truncated power-law step lengths 
distributions that allow to reproduce some of the properties extracted from the gambling logs. 
Analyzing the mean squared displacement and the first-passage time distribution for these models
allows to identify the key features needed for observing this crossover
from super-diffusion to normal diffusion.
\end{abstract}

\maketitle

\section{Introduction}
Recent years have seen a tremendous increase in online gambling, as witnessed by the emergence of numerous
online gambling sites. This surge has yielded numerous recent scientific studies, with a focus on legal,
social and psychological aspects, see \cite{Redondo15,Owens16,Gainsbury16,Goldstein16,Choliz16,Konietzny17,
Montes17,Bitar17,Hing17,Gonzalez17,Auer17,Edgren17,Martinelli17,Holden17,Sylvester17,Griffiths17} for some recent references.
In parallel to this, the quick expansion of the video gaming industry has resulted in the formation
of a huge market for virtual (in-game) item economy. Due to its easy accessibility, low entry barrier, 
and immediate outcome, virtual item gambling has become popular among game players.
In virtual item gambling, instead of directly using cash,
gamblers place bets with virtual items as virtual currencies \cite{Martinelli17,Holden17,Sylvester17}.
The virtual items here particularly refer to in-game cosmetic skins from
video games like Counter-Strike: Global Offensive, Team Fortress 2, DOTA 2, etc., 
which can be obtained through regular gameplay, in-game purchase, community market purchase, or
trading among players. Based on current estimations,
virtual item gambling industry has reached multi-billion level \cite{Grove16} and 
is expected to continue increasing.
For such a booming industry, it becomes important to be able to model the complex virtual item gambling behaviors 
at both the individual and the aggregate level. Indeed, understanding online gambling patterns is quickly becoming 
a pressing need for adolescent gambling prevention, virtual gambling regulation, and online irrationality research.

In this paper we apply the methods of statistical physics in order to develop an understanding of the
behavior of online gamblers. This is supplemented 
by the study of different random walk models that allow to recover some of the features extracted from
the empirical data. While we are not aware of any previous similar attempts to investigate online
gambling, we point out that related approaches have been used in the past in the study of horse race
betting \cite{Park01,Ichinomiya06}. More recently, online lowest unique bid auctions have been the 
subject of different studies that successfully applied the toolbox of statistical and non-linear physics 
\cite{Radicchi12a,Radicchi12b,Baronchelli13,Pigolotti12,Juul13,Zhao14}.

We focus in the following on a specific type of virtual item gambling, namely jackpot, a lottery style
game which occupies about half of the virtual item gambling market \cite{Grove16}. 
Our analysis is based on the publicly available gambling logs from a medium-sized skin gambling site
\cite{casino}. The rules of
jackpot gambling are simple: players purchase lottery tickets with skins, there will be only one winning 
ticket, and the winner takes it all. In another way of speaking, this is a parimutuel
betting type of gambling, where players place wagers in a pool, whereas
only one player is chosen as the winner and wins all the wagers in the pool. The chance of winning
equals the share of the player's wagers to the total wager pool. 

In the next Section we provide a more in-depth discussion of jackpot gambling and of the data
used in our analysis. We also discuss the models used for describing the distributions of different
quantities as well as the model selection and parameter estimation. Section III summarizes results
that we obtain from a statistical analysis of the gambling logs. 
In Section IV we view the net income of players as random walks, whereas in Section V we discuss some 
random walk models that allow to understand some of the behavioral data at the aggregate level.
We conclude in Section VI.

\section{Data and methods}

\subsection{Online jackpot game and gambling logs}
The rules of the jackpot game are very simple.
The gambling site constantly hosts a single jackpot game that any player can attend. A round can last from
a few seconds to several minutes. To take part in the game, a player needs to place a bet with lottery tickets 
purchased with one or several in-game skins deposited to the gambling site. Each ticket is usually equivalent 
to 1 US cent, and the values of the skins are calculated based on their prices listed in the community market.
There is only one winning ticket in each round of game. This  winning ticket is drawn when the total number 
of skins deposited as wagers in that round exceeds a certain threshold.
The draw is based on a uniformly distributed random number with a range equal to the total number of tickets 
purchased in that round. The player who holds the winning ticket will be the winner.
The winner wins all the wagers, which are the deposited skins in that round, after a site cut (percentage cut)
has been subtracted.

\begin{figure}
\includegraphics[width=0.6\columnwidth,clip=true]{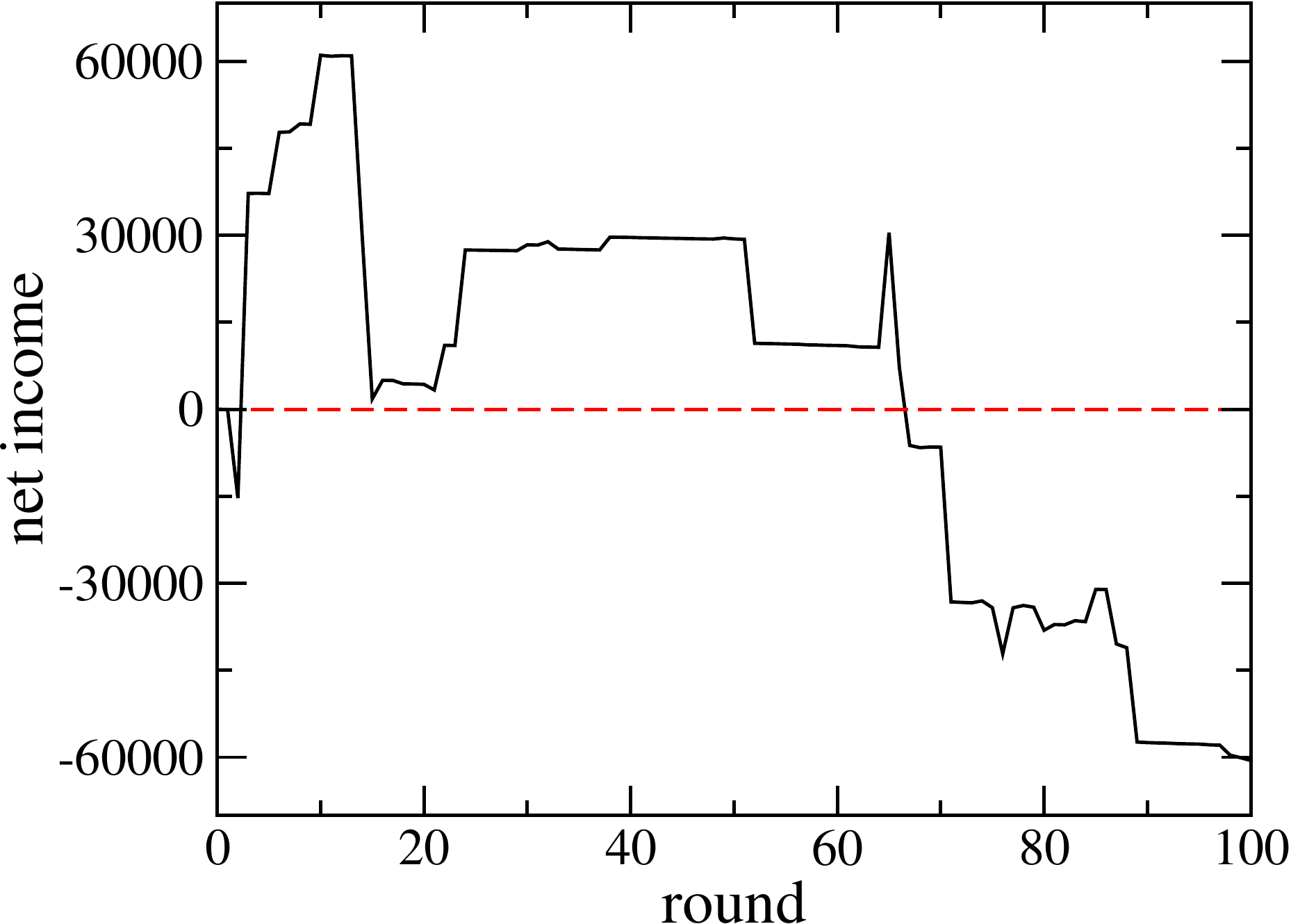}
\caption{Net income vs the number of rounds played by an online gambler. Typically, these curves
exhibit a large number of small steps and a small number of large steps.}
\label{fig1}
\end{figure}

From the rules follows that in each round a player's winning chance is determined
by the fraction their bet contributes to the total wager value of that round.
With a site cut $c$ the expected payoff $\eta$ for one player with bet value $b$ in a round with total wager $j$ is then
\begin{equation}
\displaystyle \eta = (1-c)\, j\times \frac{b}{j} - b = -c \, b,
\end{equation}
which is always negative due to the site cut. If the random number generator is well designed, then winning or 
losing a game is totally chance based, with no skill effort, similar to roulette in casinos. It is interesting to 
explore the players' gambling behaviors knowing that the expected net income is always negative. 
Fig. \ref{fig1} provides an example of the total net income for a typical gambler. The movement consists of
a large number of small steps and a few large jumps which suggest the use of a random walk based model to describe
the change of net income.

%####################### Figure 1 #############################%
%####################### Figure 1 #############################%

The publicly available gambling logs used in the following are published in the history page of the
gambling site \cite{casino}. We collected the logs of $118590$ gambling rounds, containing $943216$ bets 
placed by $105307$ players in $232$ days, from March 10, 2015, the date the site was established, to October 28, 2015.
The total wager in our study sums up to $2029835330$ tickets, which is  equivalent to about 20 million US Dollars, 
as calculated based on the players' deposited skin values. The competition is exclusively among players: 
the gambling site only takes cuts (3\% of the total wager in each round), 
but is not directly involved in gambling, except through the drawing 
of the winning tickets. In each round, the winning ticket will be drawn when there are more than 50 skins 
placed as wagers. The dataset contains information on bet ID, round index, player ID, time stamp, 
number of tickets purchased, and winner ID. Various other quantities, such as current total number and 
final total number of purchased tickets, winning chance, net gain or net loss with and without site cut, can
be calculated from these data.

The gamblers' wealth data have been collected in June 2017 from the game statistics site CS:GO BACKPACK 
\cite{csgobackpack}, which provides the gamblers' inventory values based on the item prices listed in 
the community market in June 2017. The wealth data therefore have been collected two years after the gambling 
activities. In this way we obtained information on the wealth data of $83249$ out of the $105307$ players
that gambled in the time frame given above. 

\subsection{Ethics of data analysis}
The data we analyzed in our study only contain publicly available information of gambling logs and in-game 
inventories, with no personally identifiable information included. On each dataset, we performed passive 
analysis with completely no interaction with any human subject. Before using the data, we acquired consent 
from the website administrators who host the data. We are not associated with any of those websites in any way. 
The purpose of our study is to help future researchers better understand human gambling behaviors in order to 
prevent adolescent gambling and problematic gambling.

\subsection{Distributions and fitting models}
Our analysis focuses on the probability distribution functions as well as on the
complementary cumulative distribution functions (CCDF) of various quantities extracted from the empirical data.
Whereas $P(X=x)$ is the probability that a random variable $X$ takes on the value $x$,
the corresponding complementary cumulative distribution function is given by
\begin{equation}
F(x) = 1 - P(X \leq x) = P(X > x)~.
\end{equation}

Power law distributions and their variants have been found in previous studies of very different human activities
\cite{Barabasi05,Gonzalez08,Radicchi12a,Wang17}. In online gambling quantities of interest often take on
discrete values which needs to be taken into account when selecting possible fitting models.

We consider six different fitting models in our distribution analysis. The discrete version of a power-law
distribution is given by \cite{Clauset09}
\begin{equation} \label{eq:fit1}
P_1(x) = \frac{1}{\zeta(\alpha, x_\text{min})} x^{-\alpha},
\end{equation}
with $x\ge x_\text{min}$, $\alpha>1$, and $\zeta(\cdot,\cdot)$ is the incomplete Zeta function.
Here and in the following $x$ is a positive integer value taken on by a random variable $X$.
For some data sets a fat tail is terminated by an exponential decay, which can be taken into account
by the discrete power-law distribution with exponential cut-off \cite{Wang17}
\begin{equation} \label{eq:fit2}
P_2(x) = \frac{1}{\displaystyle Li_\alpha \left(e^{-\lambda}\right) -\sum\limits_{k=1}^{x_\text{min}-1} 
k^{-\alpha} e^{-\lambda k}} x^{-\alpha} e^{-\lambda x},
\end{equation}
where $x\ge x_\text{min}$, $\lambda >0$, $\alpha>0$, and $Li_\alpha(\cdot)$ is the polylogarithm function.
Another heavy-tailed distribution is the log-normal distribution with the discrete version \cite{Wang17}
\begin{equation} \label{eq:fit3}
P_3(x) = \frac{\Phi\left(\frac{\ln(x+1)-\mu}{\sigma}\right)-\Phi\left(\frac{\ln(x)-\mu}{\sigma}\right)}{\Phi
\left(\frac{\ln(x_\text{min})-\mu}{\sigma}\right)},
\end{equation}
where $x\ge x_\text{min}$, $\sigma >0$, and $\Phi(\cdot)$ is the normal cumulative distribution. A forth basic
model is the discrete exponential function \cite{Clauset09} 
\begin{equation} \label{eq:fit4}
P_4(x) = (1-e^{-\lambda})e^{\lambda x_\text{min}} e^{-\lambda x},
\end{equation}
where $x\ge x_\text{min}$ and $\lambda >0$. Finally, we also consider two
more complex models, namely the discrete shifted power-law distribution with exponential cut-off
\begin{equation} \label{eq:fit5}
P_5(x) = C \frac{(x-\delta)^{-\alpha}}{\displaystyle 1+e^{\lambda (x-\beta)}},
\end{equation}
where $ x\ge x_\text{min}$, $\lambda > 0$, $\delta<x_\text{min}$, $\beta>x_\text{min}$,  
and $C=\displaystyle \left(\sum_{k=x_\text{min}}^\infty \frac{(k-\delta)^{-\alpha}}{1+e^{\lambda (k-\beta)}}
\right)^{-1}$ is the normalization factor, and the discrete pairwise power-law model \cite{Wang17}
\begin{equation} \label{eq:fit6}
P_6(x) = \left\lbrace \begin{split} &C \ x^{-\alpha}, \quad x_\text{min}\le x <  x_\text{trans} \\
&C x_\text{trans}^{\beta-\alpha} \ x^{-\beta}, \quad x_\text{trans}  \le x
\end{split}\right.,
\end{equation}
where $\alpha>0$, $\beta>1$, $x_\text{trans}>x_\text{min}$, and the normalizing factor 
$\displaystyle C= \left( \zeta(\alpha, x_\text{min})-\zeta(\alpha, x_\text{trans} ) + 
x_\text{trans}^{\beta-\alpha} \zeta(\beta, x_\text{trans} ) \right)^{-1}.$

We note that all these probability distributions contain a minimal value $x_\text{min}$ that defines
the range of values used for the modeling. For most quantities we choose as $x_\text{min}$ the value of $x$ that minimizes
the Kolmogorov-Smirnov statistics between the empirical and fitted distributions \cite{Clauset09}.

For a given data set we estimate for each distribution the model parameters with the maximum likelihood
method. The best fitting model is then selected using the Akaike Information Criterion (AIC). We refer the
interested reader to Appendix B in reference \cite{Wang17} for a detailed discussion.

\section{Behavioral analysis}

\subsection{Some basic statistics}

In Table \ref{basic_stat} we provide some basic statistics for the data used in our study. The huge diversity of the data 
is obvious from the very large values of the standard deviations. A meaningful analysis of the gambling data needs to consider
probability distributions (or, equivalently, complementary cumulative distribution functions).

%####################### Table 1 #############################%
\begin{table*}[!hbt]
\centering
\begin{tabular}{|l|c|c|c|c|c|}
\hline
  & mean     & minimum     & maximum      & standard deviation       & \begin{tabular}[c]{@{}r@{}}50\% percentile\end{tabular} \\ \hline
bet value                                                                    & 2309.86  & 2       & 278247   & 8429.46  & 91 \\ \hline
total net income                                                                    & $-578.88$  & $-773524$ & 751635   & 15513.36 & $-150$ \\
number of rounds a player attended & 8.34     & 1       & 1931     & 31.94    & 2 \\ \hline
number of players in a round                                                        & 7.41     & 1       & 25       & 2.15     & 7 \\
jackpot value                                                                       & 17116.41 & 100     & 396760   & 24399.50  & 7548 \\ \hline
\end{tabular}
\caption{Basic statistics for the gambling data used in this study.}
\label{basic_stat}
\end{table*}
%####################### Table 1 #############################%

\subsection{Distributions}

%####################### Figure 2 #############################%
\begin{figure}
\includegraphics[width=0.6\columnwidth,clip=true]{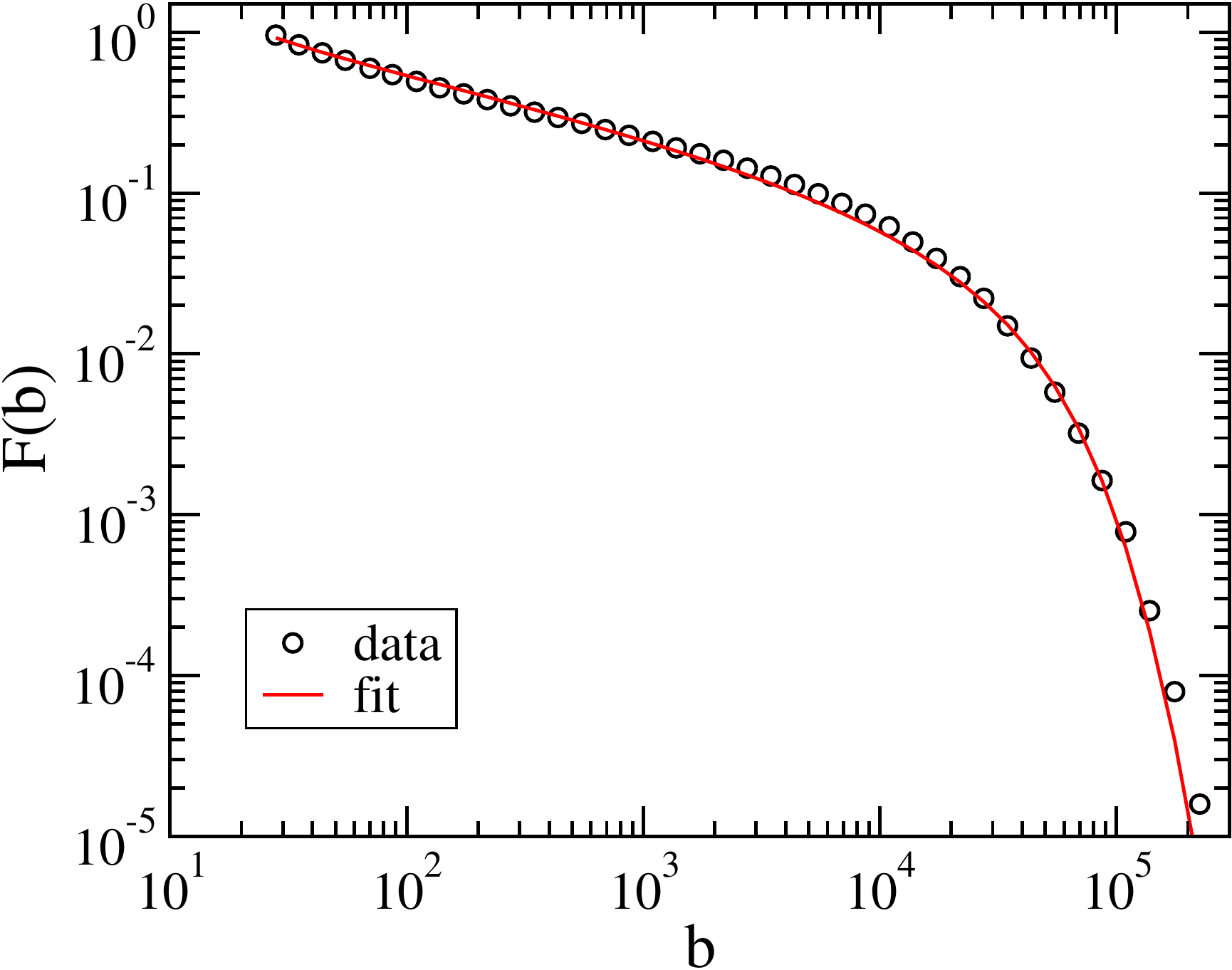}
\caption{The complementary cumulative distribution function for bet values. The best fit is obtained for
a shifted power law with an exponential cutoff, see Eq. (\ref{eq:fit5}), with $b_{min}=25$ and the maximum likelihood estimators
$\alpha = 1.297$, $\lambda = 3.429 \times 10^{-5}$, $\delta = 9.905$, and $\beta = 4.629 \times 10^4$.}
\label{fig2}
\end{figure}
%####################### Figure 2 #############################%

%####################### Figure 3 #############################%
\begin{figure}
\includegraphics[width=0.7\columnwidth,clip=true]{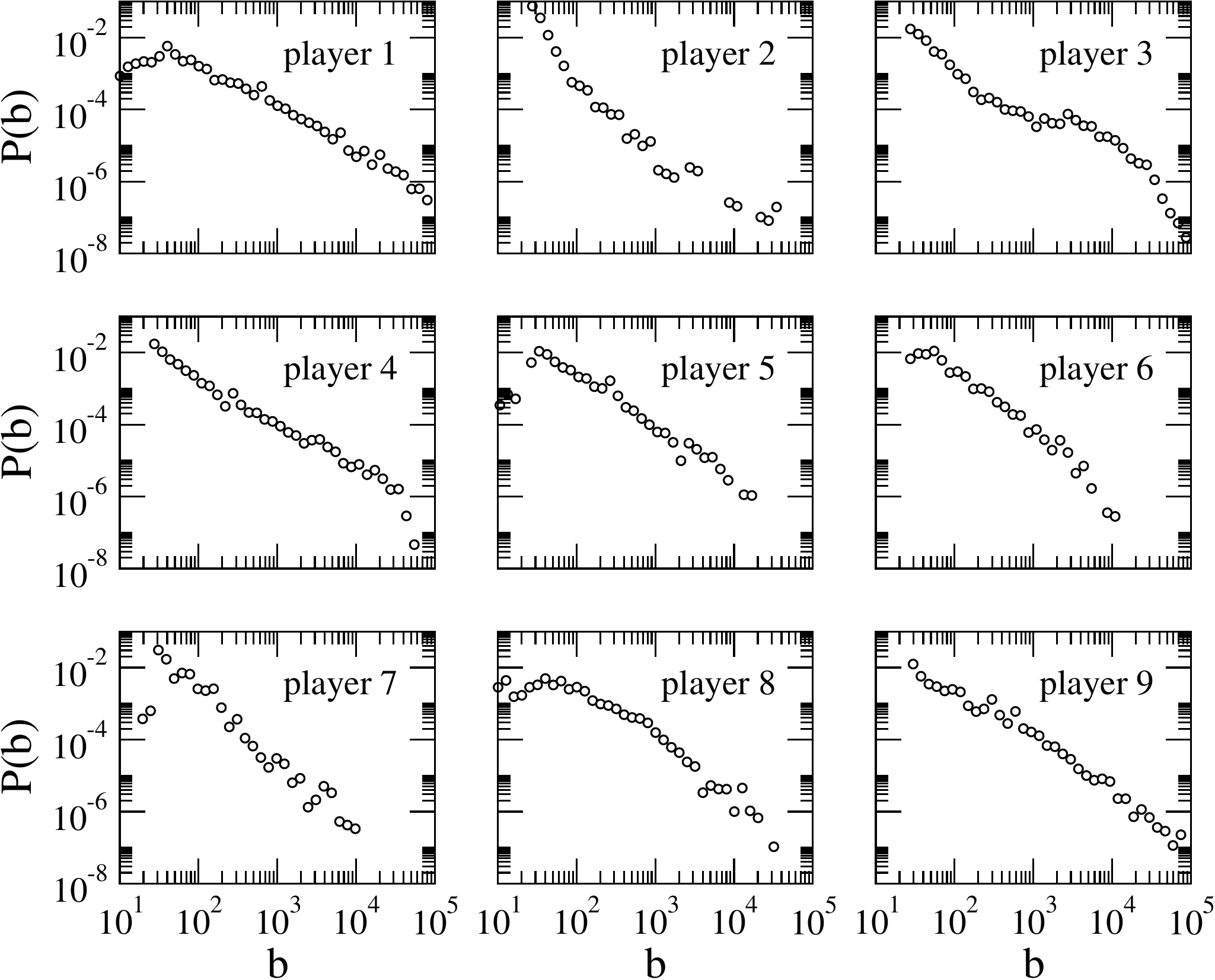}
\caption{Wager probability distributions for the nine players with the largest numbers of bets
(ranging from 1931 bets for player 1 to 1286 for player 9). Heavy tails are present in all nine
distributions.
}
\label{fig3}
\end{figure}
%####################### Figure 3 #############################%

A fundamental quantity for our analysis is the bet value, and the distribution of bet values allows one to gain a quick understanding
of betting patterns. As shown in Fig. \ref{fig2}, the complementary cumulative distribution function for the bet value 
at the aggregate level is
described by a shifted power law with an exponential cutoff: bet values smaller than $\beta \sim 4.6 \times 10^4$ 
follow a power-law distribution, whereas very large bets are distributed exponentially (such guaranteeing a finite variance). 
The heavy-tail property of the bet distribution is also readily identified when studying the bet value distributions
of individual gamblers. Fig. \ref{fig3} shows the wager distribution for the nine players which played the largest numbers of rounds
(between 1931 and 1286). While there is some variability in these distributions, they all exhibit heavy tails in the form
of power laws with exponents typically in the range $\left[ 1.1, 1.7 \right]$.

%####################### Figure 4 #############################%
\begin{figure}
\includegraphics[width=0.6\columnwidth,clip=true]{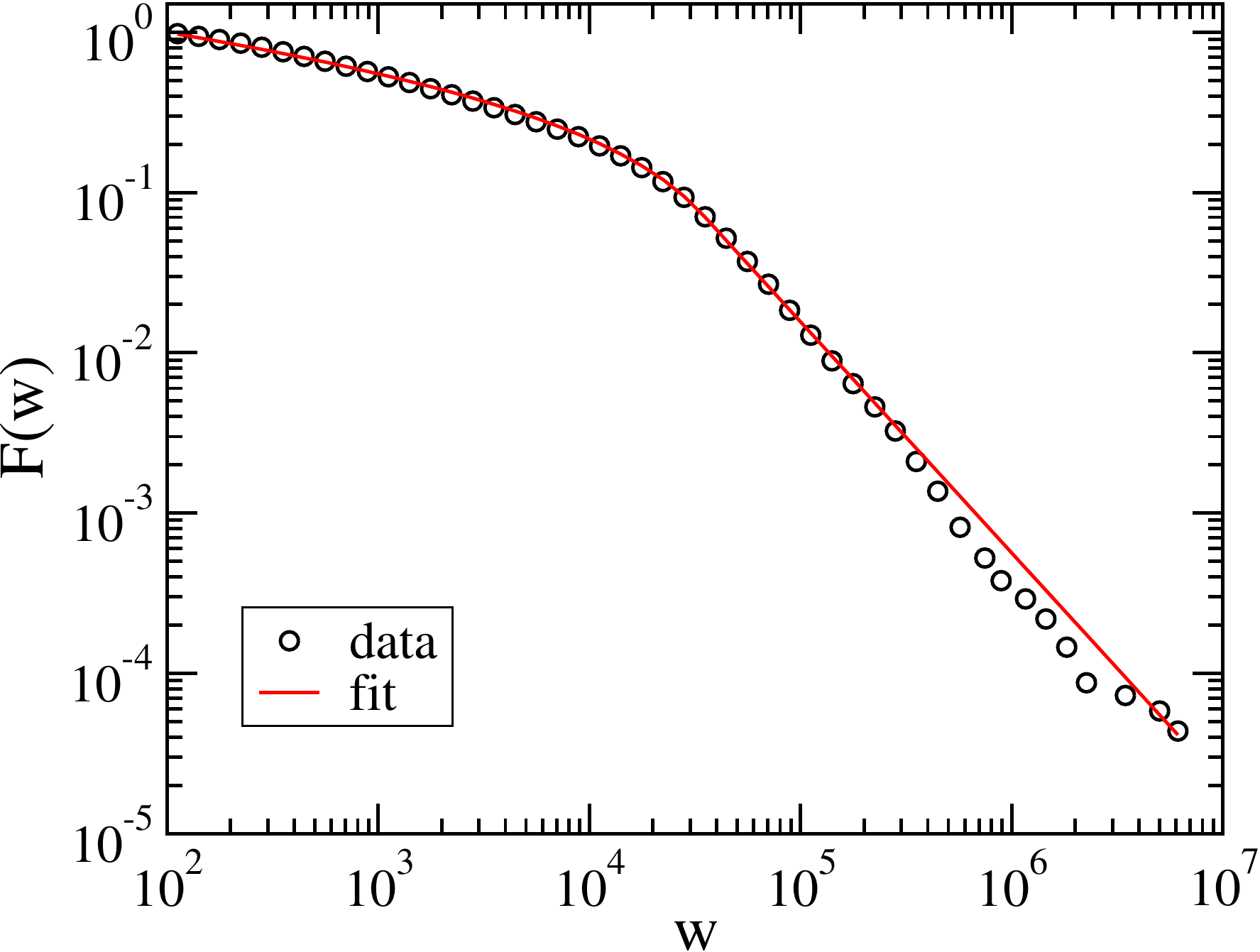}
\caption{The complementary cumulative distribution function of the gambler's wealth $w$, where one unit corresponds to 1 US Cent.
These data have been collected in June 2017 from the game statistics site CS:GO BACKPACK
\cite{csgobackpack}. The best fit of the data is achieved with a pairwise power-law
distribution (\ref{eq:fit6}) with the maximum likelihood estimator $\alpha = 1.128$ and
$\beta=2.442$ as well as with the parameters $w_\text{min}=100$ and $w_\text{trans}=33928$.}
\label{fig4}
\end{figure}
%####################### Figure 4 #############################%

In gambling a player's wealth provides a natural upper limit for possible bet values. Studies have shown that
the net wealth distribution in human society follows a distribution that combines an exponential decay for small values
and a power-law tail for large values \cite{Yakovenko09}. For the online gamblers' wealth, this is different, see
Fig. \ref{fig4}. We still have a power-law tail for large values (with an exponent $\beta = 2.442$), but for small values
the exponential decay is replaced by a power-law decay with an exponent $\alpha = 1.128$. For this figure we computed
the wealth of each player by taking the sum of the values (community market price) of the skins in each player's
inventory.

%####################### Figure 5 #############################%
\begin{figure}
\includegraphics[width=0.6\columnwidth,clip=true]{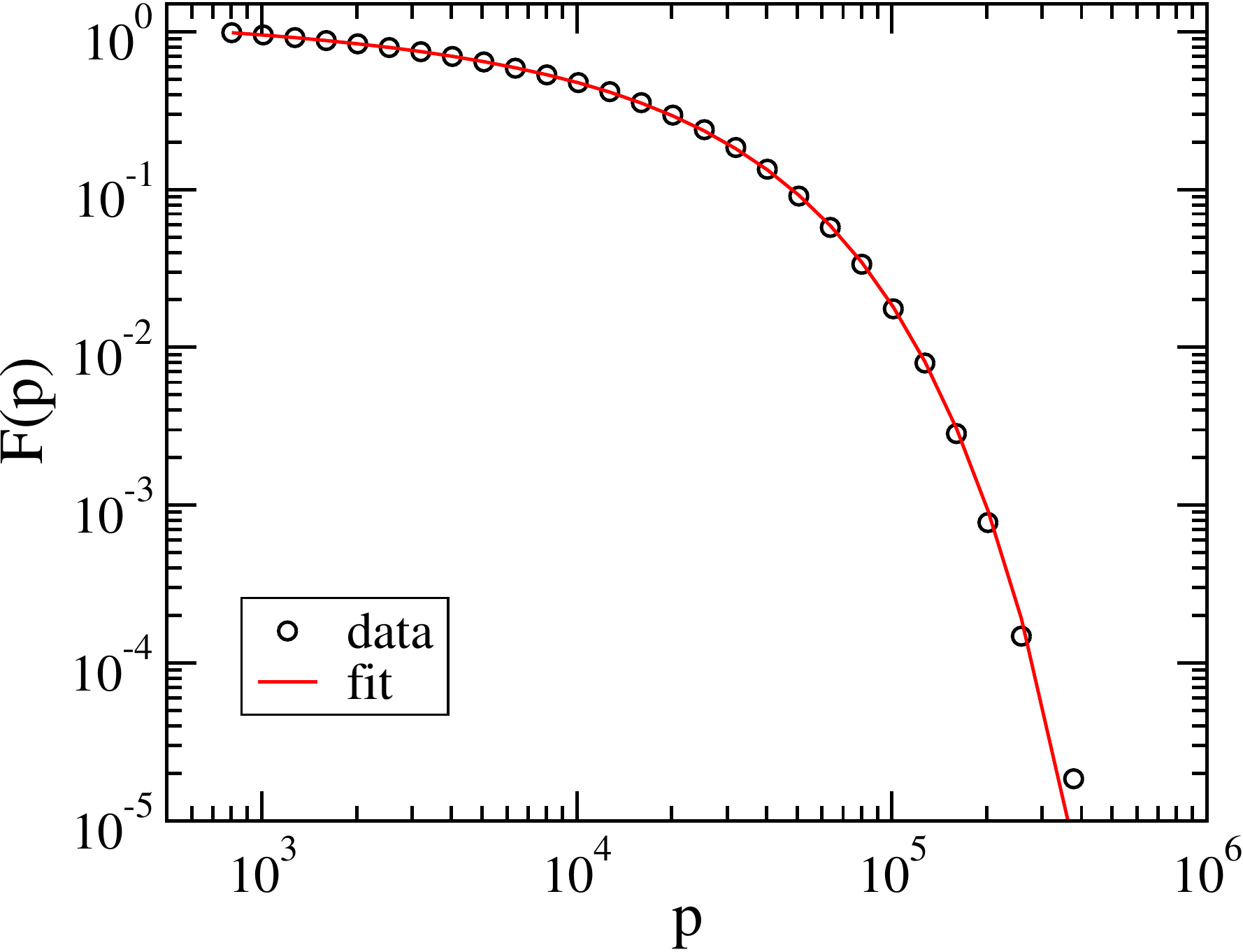}
\caption{The complementary cumulative distribution function of the pool size (i.e. the total wager in one round) $p$.
The fitting curve is a power-law with exponential cut-off (\ref{eq:fit2}) with the maximum likelihood estimators
$\alpha = 0.650$ and $\lambda = 2.577 \times 10^{-5}$.}
\label{fig5}
\end{figure}
%####################### Figure 5 #############################%

In each round a gambler either loses their wager or wins the whole pool (minus the site cut), resulting
in the random walk like behavior of the net income shown in Fig. \ref{fig1}. The probability distribution
of the pool size is described by a power-law distribution with an exponent $a=0.650$ that
ends in an exponential cut-off, see Fig. \ref{fig5}. The same functional form is
found if we consider the wins instead of the pool sizes (see Section IV).

%####################### Figure 6 #############################%
\begin{figure}
\includegraphics[width=0.6\columnwidth,clip=true]{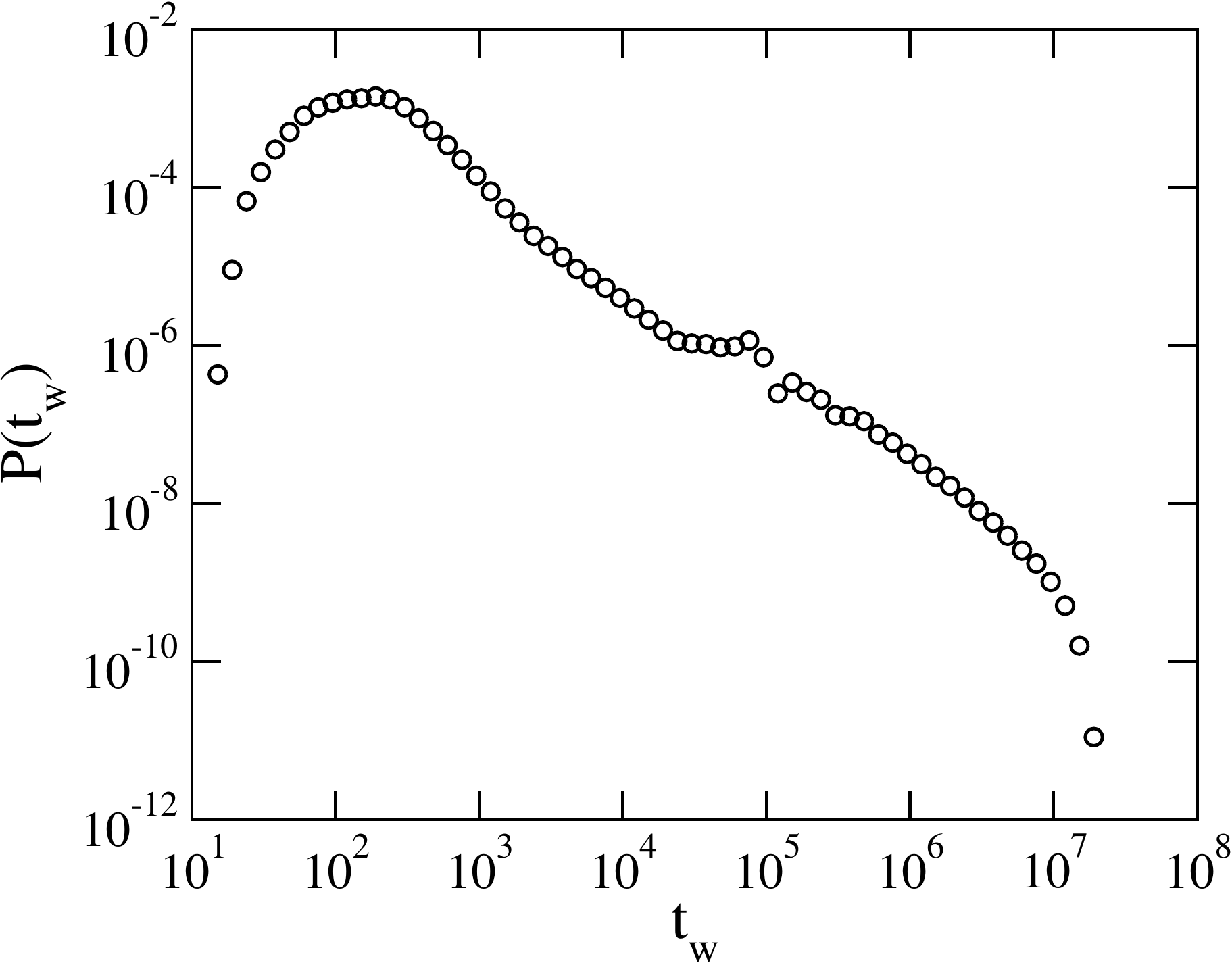}
\caption{The probability distribution of the waiting time between successive gambles. The waiting
time is measured in seconds.
}
\label{fig6}
\end{figure}
%####################### Figure 6 #############################%

%####################### Figure 7 #############################%
\begin{figure}
\includegraphics[width=0.6\columnwidth,clip=true]{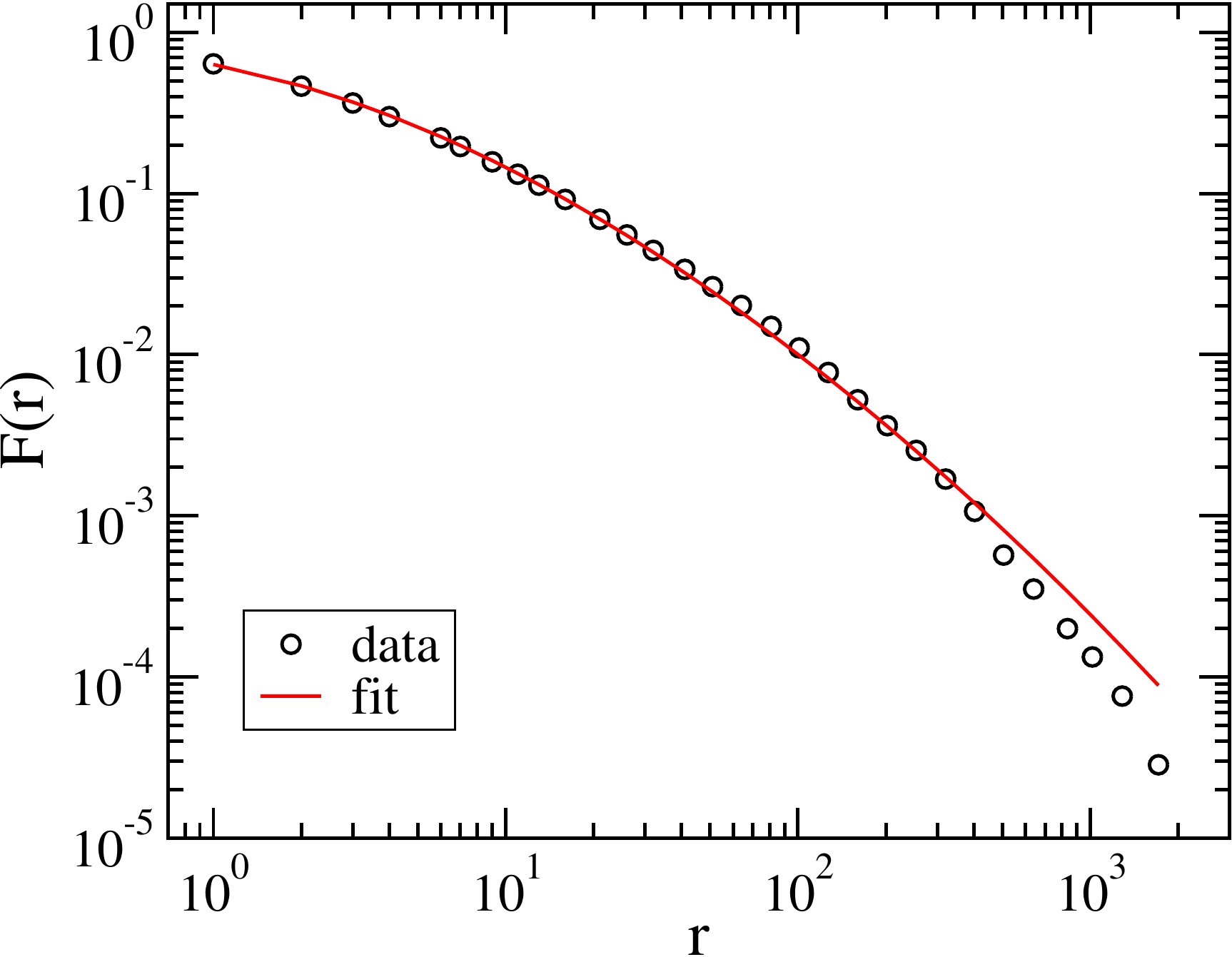}
\caption{
The complementary cumulative distribution function of the number of rounds $r$ played by individual players.
The data are best fitted by a log-normal distribution (\ref{eq:fit3}) with the maximum likelihood estimators
$\mu = -1.777$ and $\sigma = 2.238$.
}
\label{fig7}
\end{figure}
%####################### Figure 7 #############################%

The available logs also allow to discuss time dependent quantities, as for example
the waiting time $t_w$, defined as the time measured in seconds between successive bets by the same user,
or the number of rounds $r$ played by individual gamblers. The waiting time probability distribution shown in
Fig. \ref{fig6} has some interesting features. The plateau for $P(t_w)$ close to  $t_w=10^5$
indicates that a sizeable portion of gamblers play bets day after day (24 hours correspond to 86,400 
seconds). The heavy tail of the distribution reveals that some persons restart gambling after a
month-long hiatus, which illustrates some of the challenges gambling prevention faces.
Fig. \ref{fig7} shows that the number of rounds played by individual players during the 232 days 
covered by the gambling logs 
is well described by a log-normal distribution. Remarkably, a sizeable number of gamblers placed thousand
and more bets during the time frame covered by the logs.

%####################### Figure 8 #############################%
\begin{figure}
\includegraphics[width=0.6\columnwidth,clip=true]{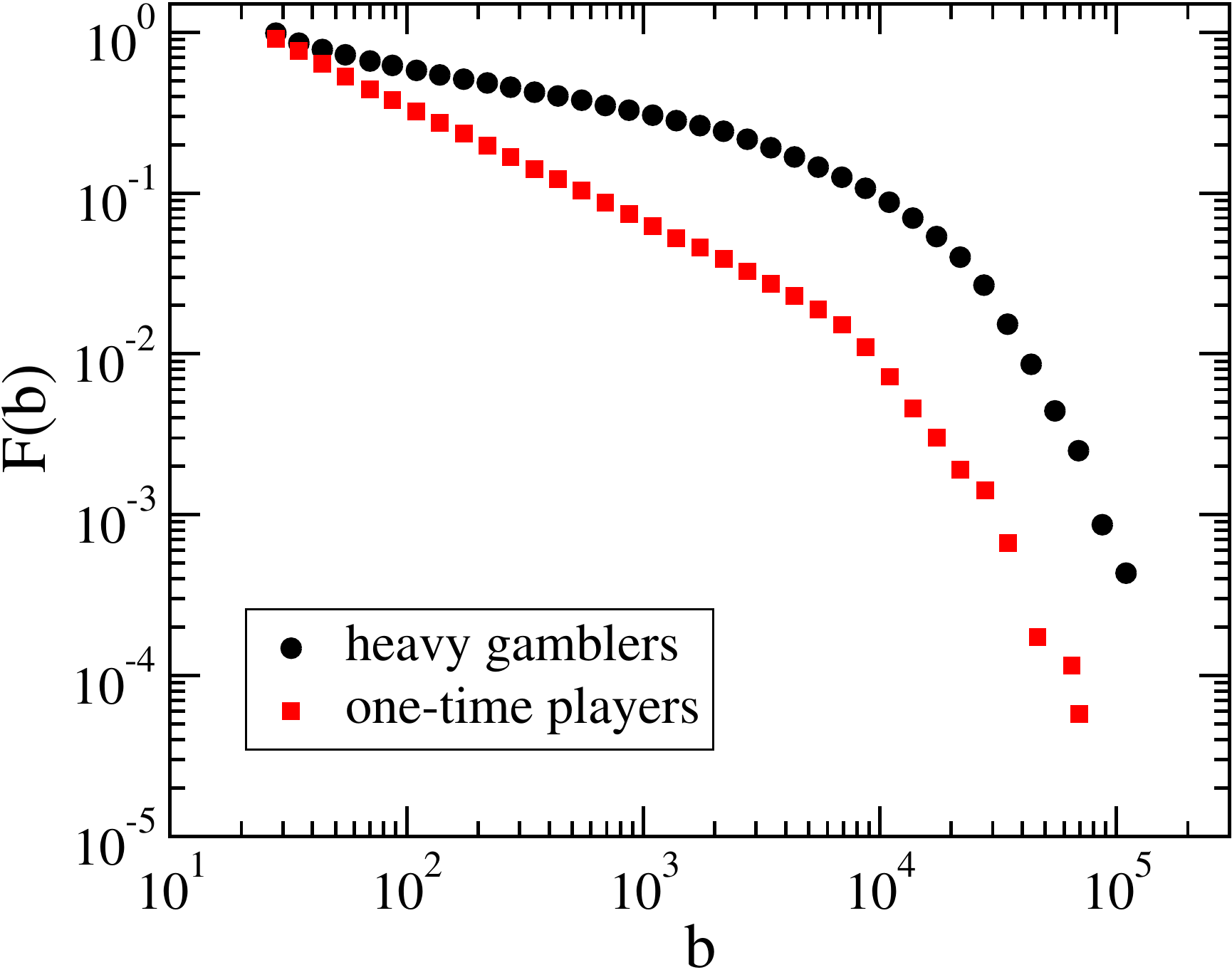}
\caption{Comparison of the betting patterns of heavy gamblers and one-time players.
Shown is the complementary cumulative distribution function for bet values.
}
\label{fig8}
\end{figure}
%####################### Figure 8 #############################%

Besides discussing data at the population level, we can also identify different sub-groups of
gamblers and discuss differences between these groups. Fig. \ref{fig8} provides one example
where we confront the distribution of bets of one-time players with that of heavy gamblers
(defined as having played at least 600 rounds). Obviously one-time players are much more
risk-averse and are therefore unlikely to bet large amounts. 

\subsection{Correlations}

Correlation coefficients help to understand the relationships between the different quantities.
As our quantities, be it outcomes, bets, and profits, all follow heavy-tailed distributions,
the standard Pearson's product-moment correlation coefficient may provide erroneous results.
More appropriate are rank-based correlation coefficients, such as Kendall's tau \cite{Kendall90}
or Spearman's rho \cite{Taylor87}. We verified that the same conclusions are obtained
from these two coefficients. For that reason we will only discuss Kendall's tau in the following.
Assuming a set of observations $\left\{ \left( x_i, y_i \right) \right\}$ of two joint variables
$x$ and $y$, Kendall's tau can be calculated as 
\begin{equation}
\tau_K(x,y) = \frac{\displaystyle \sum\limits_{i<j} {\rm sgn} \left[(x_i-x_j)(y_i-y_j) \right]}{\displaystyle \sqrt{\frac{1}{2}
n(n-1) - U} \sqrt{\frac{1}{2}n(n-1) - V}}
\end{equation}
where $\rm{sgn}$ is the signum function, whereas $U$ and $V$ are the numbers of $x$-tied pairs and $y$-tied pairs.

For each player the gambling history can be summarized as a sequence $\left\{ \left( b_i, o_i \right) \right\}$
where $b_i$ is the value of the $i$-th bet and $o_i$ is the outcome of that round. When losing the round,
then the outcome is the negative of the bet value, whereas for a winning round $o_i$ is the total bet value
minus the winner's wager and the site cut. Focusing on the 2,318 players that attended more than 60 rounds, we can
obtain from these data different correlation coefficients.

%####################### Figure 9 #############################%
\begin{figure}
\includegraphics[width=0.8\columnwidth,clip=true]{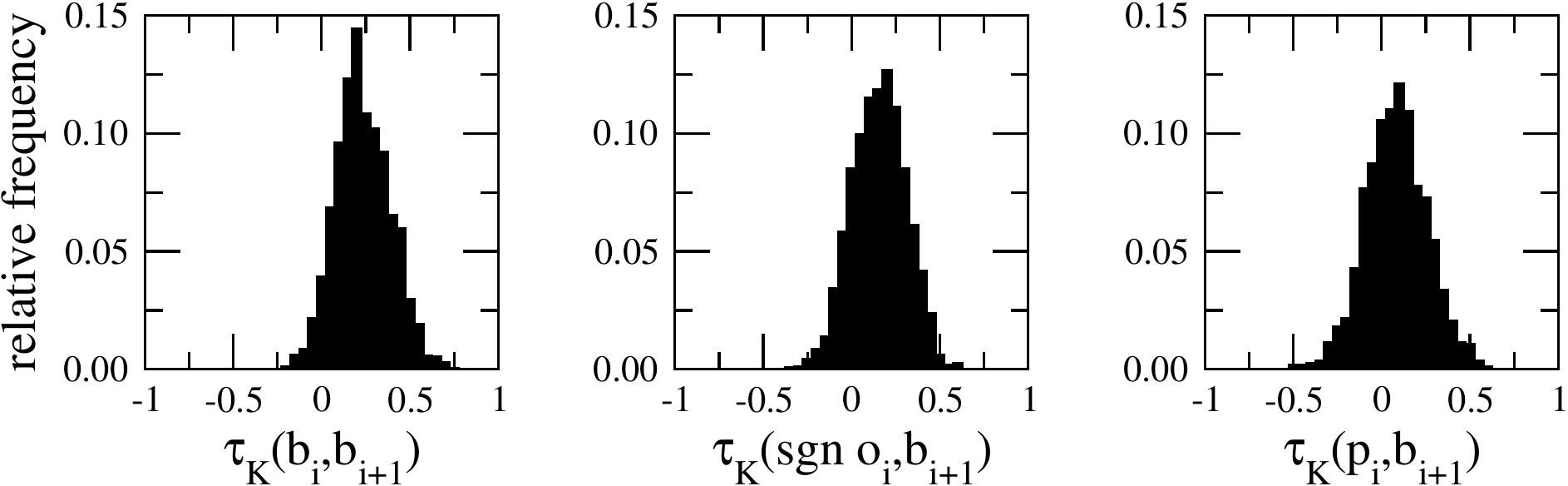}
\caption{The relative frequencies for the three correlation coefficients discussed in the text.
Left panel: correlation between successive bets, with the mean value 0.260. 
Center panel: correlation between the sign of a bet outcome and the next bet, with the mean value 0.181.
Right panel: correlation between the profit and the subsequent bet, with the mean value 0.107.
}
\label{fig9}
\end{figure}
%####################### Figure 9 #############################%

The correlation between successive bets $\tau_K(b_i,b_{i+1})$ is positive for most players, with an average value
$\tau_K = 0.260$. The relative frequency of a given value of $\tau_K(b_i,b_{i+1})$ is displayed in the
left panel of Fig. \ref{fig9}. In order to understand this graph we remark that a negative value is obtained when
a gambler places larger and smaller bets in turn, whereas placing bets randomly yields a value close to zero. From the graph
follows that only few gamblers have these types of gambling behavior. Instead, for most gamblers 
bets are not independent but indicate some level of memory. Indeed, positive correlation indicates a consistent
betting behavior without dramatic changes from bet to bet.

Also shown in Fig. \ref{fig9} are the relative frequencies for the correlation between the sign of a 
bet outcome and the next bet, $\tau_K(\rm{sgn}~o_i,b_{i+1})$, and the correlation between the profit $p_i$ (i.e. the value
of the outcome in case it is positive) and the subsequent bet, $\tau_K(p_i,b_{i+1})$. The first correlation coefficient helps to
understand how gaining/losing money affects the next bet, whereas the second one shows whether a bet value is affected
by the value of the previous profit. Profit corresponds to positive outcome, so that for 
the computation of $\tau_K(p_i,b_{i+1})$ we remove
all bets with a negative outcome $o_i$. For both correlations we restrict ourselves to players who made profit in at least 15
rounds and had negative outcomes in also at least 15 rounds. This yields 1,608 eligible players. The relative frequencies
shown in the center and right panels of
Fig. \ref{fig9} reveal for most players a weak positive correlation between the betting value and the outcome/profit.
There is a tendency for gamblers to place larger respectively smaller bets in case the outcome in the preceding round was positive respectively
negative.

\section{Net income viewed as a random walk}

As we have already seen in Fig. \ref{fig1}, the net income of a player changes at each round where they place a bet,
due to winning or losing that round. This then generates a time series where ``time'' is increased by one at each
round played by the gambler and suggests a description as a random walk in the one-dimensional space of net income.
Of course the random walkers are not independent as the loss of one gambler will be part of the gain of another one.
Also, the fact that every gambler has a finite wealth will put constraints on the random walk.

%####################### Figure 10 #############################%
\begin{figure}
\includegraphics[width=0.6\columnwidth,clip=true]{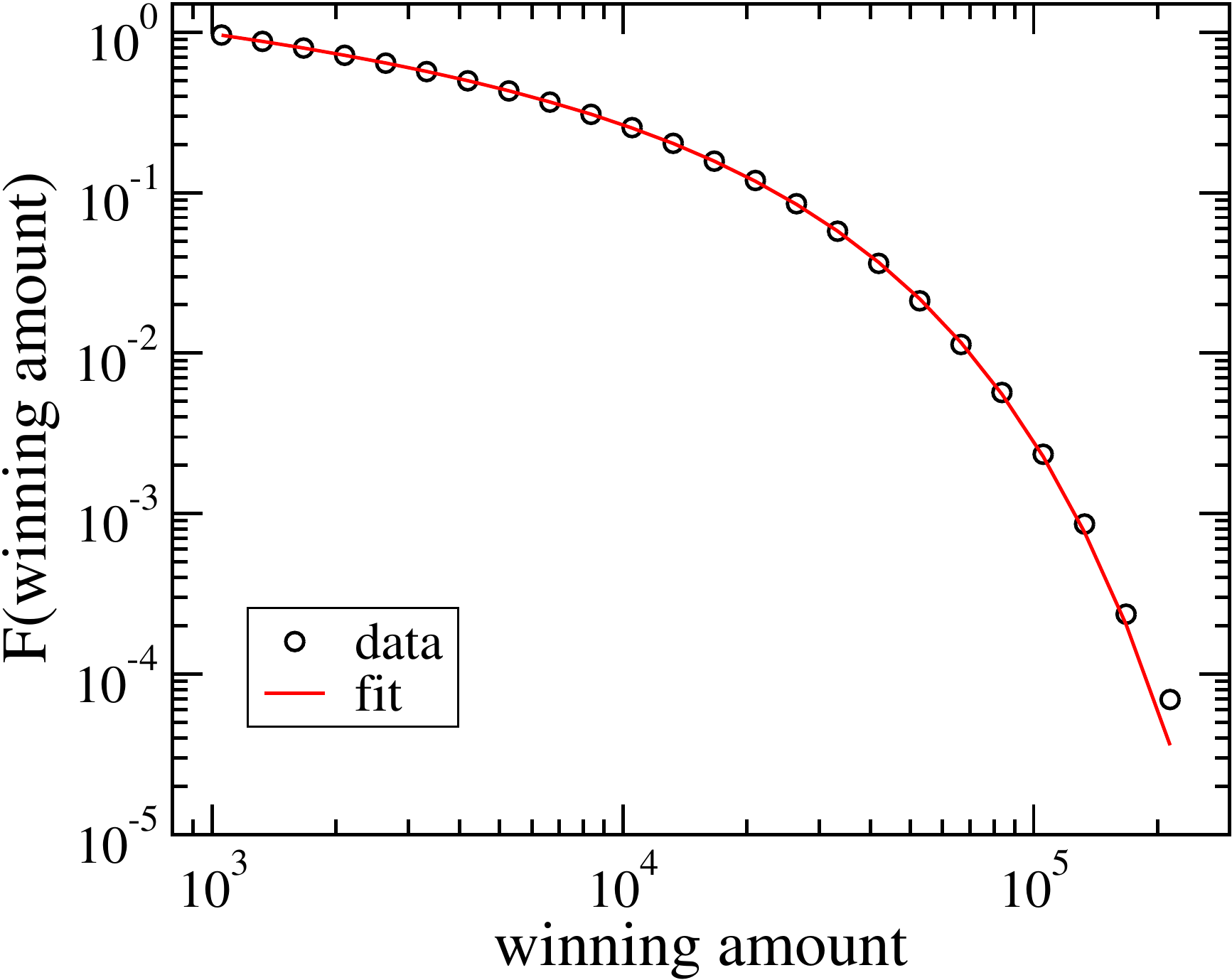}
\caption{The complementary cumulative distribution function of the winning amounts.
The fitting curve is a power-law with exponential cut-off (\ref{eq:fit2}) with the maximum likelihood estimators
$\alpha = 1.063$ and $\lambda = 3.192 \times 10^{-5}$.}
\label{fig10}
\end{figure}
%####################### Figure 10 #############################%

The jumps done by our random walkers have the peculiarity that they follow different distributions depending
on whether they jump ``left'' (net income decreases after losing a round) or ``right'' (net income increases 
after winning a round). ``Left'' and ``right'' indicate the relative decrease or increase with respect to the value
of the net income before the round is played. The distribution of losses is very similar to the distribution of
bet values (as in a given round all bets result in losses with the exception of the winning bet). 
As shown in Fig. \ref{fig2}, this distribution is described at the aggregate level by a shifted 
power law with an exponential cutoff. Power laws are also observed in Fig. 3 for individual gamblers.
The distribution of winning amounts shown in Fig. \ref{fig10} is well described by a 
power-law distribution with an exponential cutoff, albeit
with a different power-law exponent $\alpha$. The fact that the distributions for jumps in both directions,
albeit not identical, are power-law distributions indicates that the random walk of the net income should follow
a truncated L\'{e}vy flight pattern.

%####################### Figure 11 #############################%
\begin{figure}
\includegraphics[width=0.6\columnwidth,clip=true]{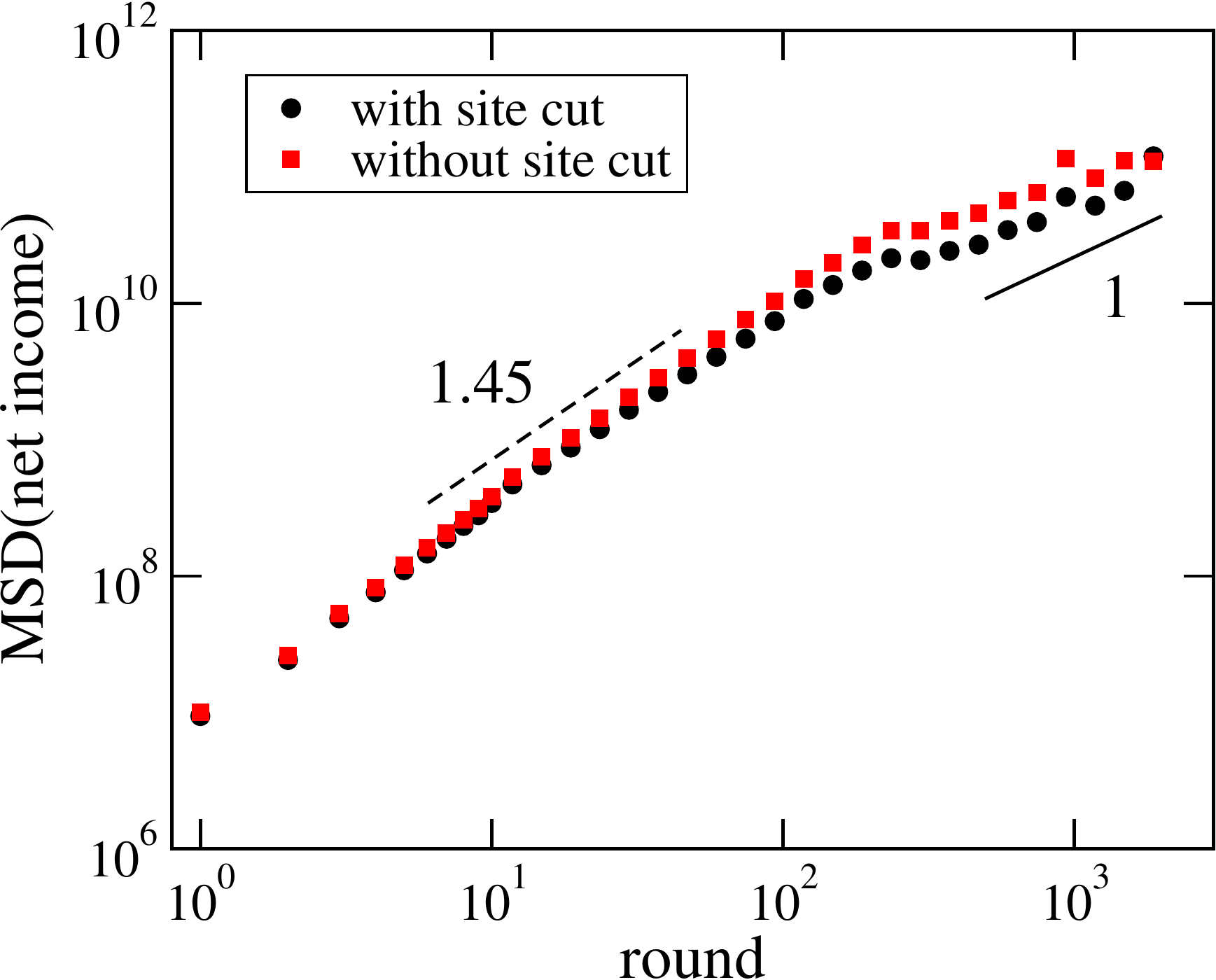}
\caption{Mean squared displacement when viewing the net income of the gamblers as a random walk, with
time measured in numbers of rounds played.
Independent on whether the site cut is considered or not, two different regimes are observed,
with the early one being super-diffusive with an exponent close to 1.45, whereas the later one
is close to normal diffusion.
}
\label{fig11}
\end{figure}
%####################### Figure 11 #############################%

Fig. \ref{fig11} shows that the mean squared displacement of the net income random walk displays a first regime
that is super-diffusive with an exponent close to 1.45. We show two curves in that figure, one where we consider as
winning amount the total pool size in a round and one where we subtract the site cut and take the remaining amount
as the length of the jump. At very late times this first regime goes over into a normal diffusion  regime,
with the measured slope close to 1 in the log-log plot. This crossover from super-diffusion to normal diffusion
is in fact expected for truncated power-law distributions \cite{Mantegna94,Inoue07} and has been observed in a variety
of systems (see, e.g., \cite{Ito03,Maruyama03,Stauffer07}).

%####################### Figure 12 #############################%
\begin{figure}
\includegraphics[width=0.6\columnwidth,clip=true]{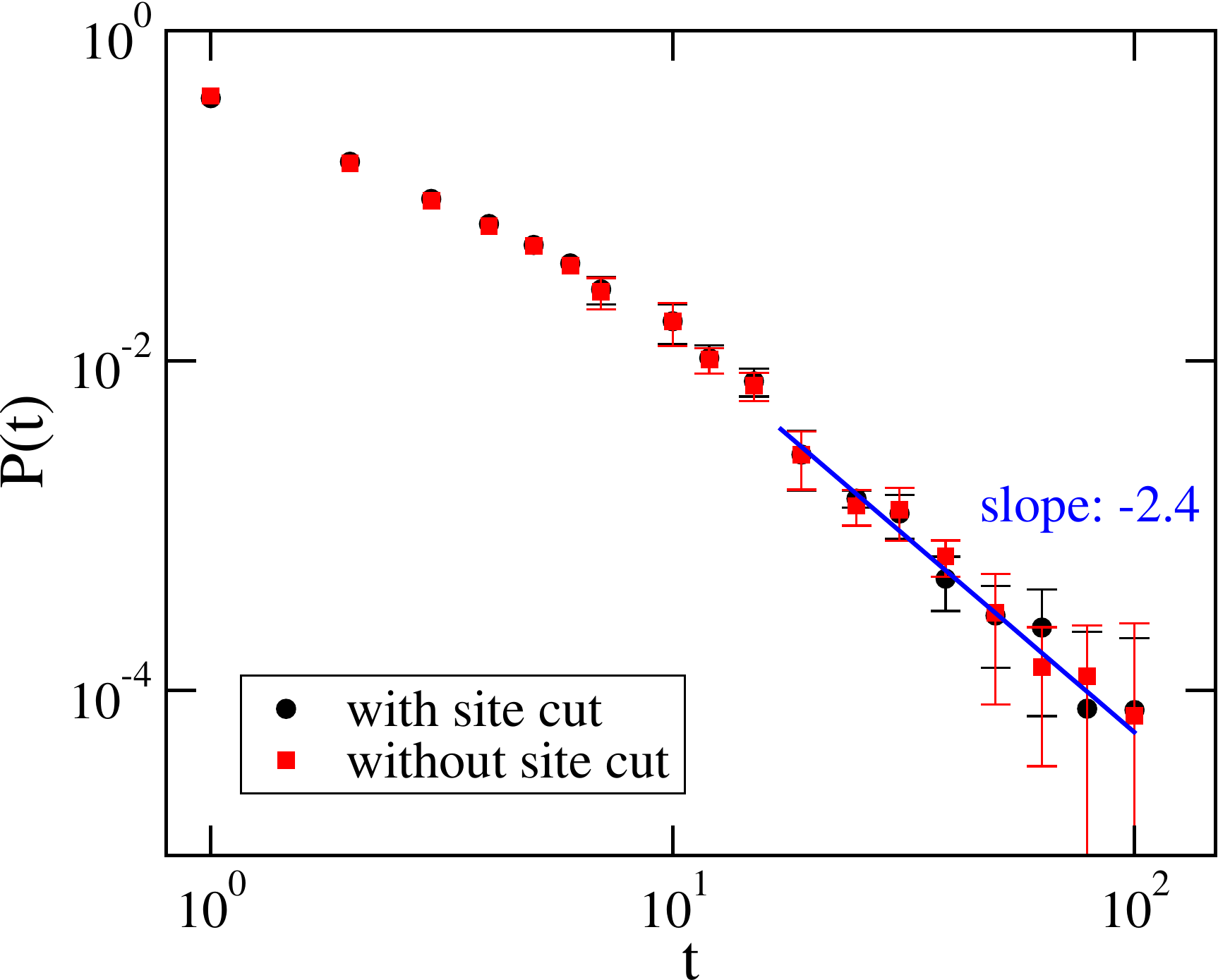}
\caption{First-passage time distribution obtained 
from the data of 387 players that gambled in more than 200 rounds. The super-diffusive
regime is revealed by a power-law decay with an exponent larger than 3/2. 
Error bars result from log-binning averaging and indicate 95\% confidence intervals.
}
\label{fig12}
\end{figure}
%####################### Figure 12 #############################%

A quantity of much interest is the first-passage time \cite{Redner01}, i.e. the time needed for a stochastic 
variable (in our case the net income viewed as a random walker) to take on for the first time a given value.
Indeed, the first-passage time distribution can help to determine the diffusive behavior of a stochastic
process \cite{Mantegna94,Inoue07}. For our stochastic process $N_r$, $r=1, \cdots, R$, representing the net income with
$R$ being the maximum number of rounds played, the first-passage time is defined by $t = \mbox{min}
\left\{r > r_0;X_k = \pm N_{fp} \right\}$, where $N_{fp}$ is the target value. As shown in \cite{Inoue07},
the first passage time distribution $P(t)$, defined as the survival probability that, starting from $r = r_0$,
the series $N_r$ stays within the range $\left[ N_{r_0} - N_{fp}, N_{r_0} + N_{fp} \right]$ up to the round $r=r_0+t$,
is given by the expression
\begin{equation} \label{fpt}
P(t) = \lim\limits_{R \longrightarrow \infty} \frac{1}{R} \sum\limits_{r=1}^R \Theta \left( \left| 
N_{r+t} - N_r \right| - N_{fp} \right) - \lim\limits_{R \longrightarrow \infty} \frac{1}{R} \sum\limits_{r=1}^R \Theta \left( \left|
N_{r+t-1} - N_r \right| - N_{fp} \right)~,
\end{equation}
where $\Theta(x)$ is the Heaviside step function. 

As Eq. (\ref{fpt}) requires sufficiently long time series, we focus on the 387 players that played at least 200 rounds
and choose $N_{fp}=500$.
The resulting first-passage time distribution is still very noisy. In order to reduce the noise we use the log-binning technique which
yields the distribution shown in Fig. \ref{fig12}. Inspection of that figure reveals that after some initial time regime a
super-diffusive regime prevails, as indicated by a slope larger than 3/2, the characteristic value for a Gaussian process.
As already mentioned, for any truncated heavy-tail distribution the long-time behavior should be normal diffusion, and we do observe
the crossover from a super-diffusive to a normal diffusive behavior in Fig. \ref{fig11} for the mean squared displacement. For the
first-passage time distribution obtained from the gambling logs the long-time normal diffusion decay with an exponent 3/2 is not readily
observed, due to the shortness of the available time series.

\section{Modeling online gambling through random walk models}

In order to better understand this switch from a super-diffusive to a normal diffusive behavior in the net income
random walk we discuss in the following three different random walk models.
The aim of this investigation is not so much to find the best parameter sets to reproduce the empirical data,
but instead to gain insights into the necessary ingredients to obtain from these models data with qualitative
similar properties as those derived from the gambling logs.

In all three models we consider that at each round four players interact (this is mostly useful for the
numerical simulations; the analytical results for the simpler models are valid for any number of gamblers interacting
in a round). For each round the gamblers place a bet
with a value taken from the continuous power law distribution with exponential cut-off
\begin{equation} 
P(b) = \frac{\lambda^{1-\alpha}}{\Gamma \left( 1-\alpha,\lambda b_{min} \right)} b^{-\alpha} e^{-\lambda b}
\label{cont_power_exp}
\end{equation}
with $\lambda > 0$ and $b \geq b_{min}$ (we choose $b_{min}=1$), whereas $\Gamma \left( \cdot,\cdot \right)$
is the incomplete Gamma function. This distribution (\ref{cont_power_exp}), motivated by the data from the gambling logs
that show a power-law behavior with an exponential cut-off, is the continuous version of
the discrete distribution (\ref{eq:fit2}).
For the results discussed in the following
we fix the mean $\left< b \right>=100$.
The two parameters $\alpha$ and $\lambda$ are then not independent
but related through that mean bet value as $\left< b \right> = \frac{ \Gamma \left( 2 - \alpha , \lambda \right)}{
\lambda \Gamma \left( 1 - \alpha, \lambda \right)}$.
We vary $\alpha$ between 1.2 and 1.6. We verified that qualitatively our results are unchanged if instead of
using the distribution (\ref{cont_power_exp}) we use a power-law distribution with a sharp truncation:
\begin{equation} \label{cont_power}
P(b)= C\, b^{-\alpha}
\end{equation} with $b \in \left[ b_{min}, b_{max} \right]$ and $C = \frac{\alpha-1}{b_{min}^{1 -\alpha}-
b_{max}^{1-\alpha}}$ where $b_{max}$ and $\alpha$ are related when fixing the
mean bet value.

Our first two models are focusing on a single gambler with infinite wealth. In model 1 \cite{footnote} we fix the winning chance of this 
gambler to be 1/4 (in a generalization to $n$ interacting gamblers, the winning chance would be $1/n$). This model does
not take into account that in the online game the winning chance is proportional to the bet value. We therefore consider 
a more realistic model 2 which implements this relationship between the bet value and the winning chance. Model 3, finally,
is a more sophisticated version of model 2 where, similarly to the online game, a large pool of gamblers is available  
(the data shown below have been obtained for $N=1,000,000$) and at each round $n=4$
gamblers are selected randomly to play the round. We calculate quantities for all players, which are no longer independent,
in contrast to models 1 and 2, and after each round we update the net income of all 4 players involved in that round.
We also take into account in model 3 that the wealth of each player is finite: before the first round is played every gambler is assigned a wealth
taken from the power-law probability distribution
\begin{equation} \label{wealth}
P(w) = \frac{1}{w_{min}} \left( \frac{w}{w_{min}} \right)^{-2}
\end{equation}
with $w_{min} = 1$. 
As in model 3 the gamblers have been provided with a finite wealth, the individual
net income random walks all have an absorbing state of zero wealth. As soon as the wealth of a gambler is zero,
this gambler is removed from the pool. While it is tempting to discuss our random walkers in the
context of previous studies of random walk type motion with absorbing boundaries \cite{Buldyrev01,Kantor07,Novikov11,Dybiec17},
it is crucial to realize that our random walkers are not independent, but instead at each round the winner's step length
is correlated to those of the losers. 

%####################### Figure 13 #############################%
\begin{figure}
\includegraphics[width=0.4\columnwidth,clip=true]{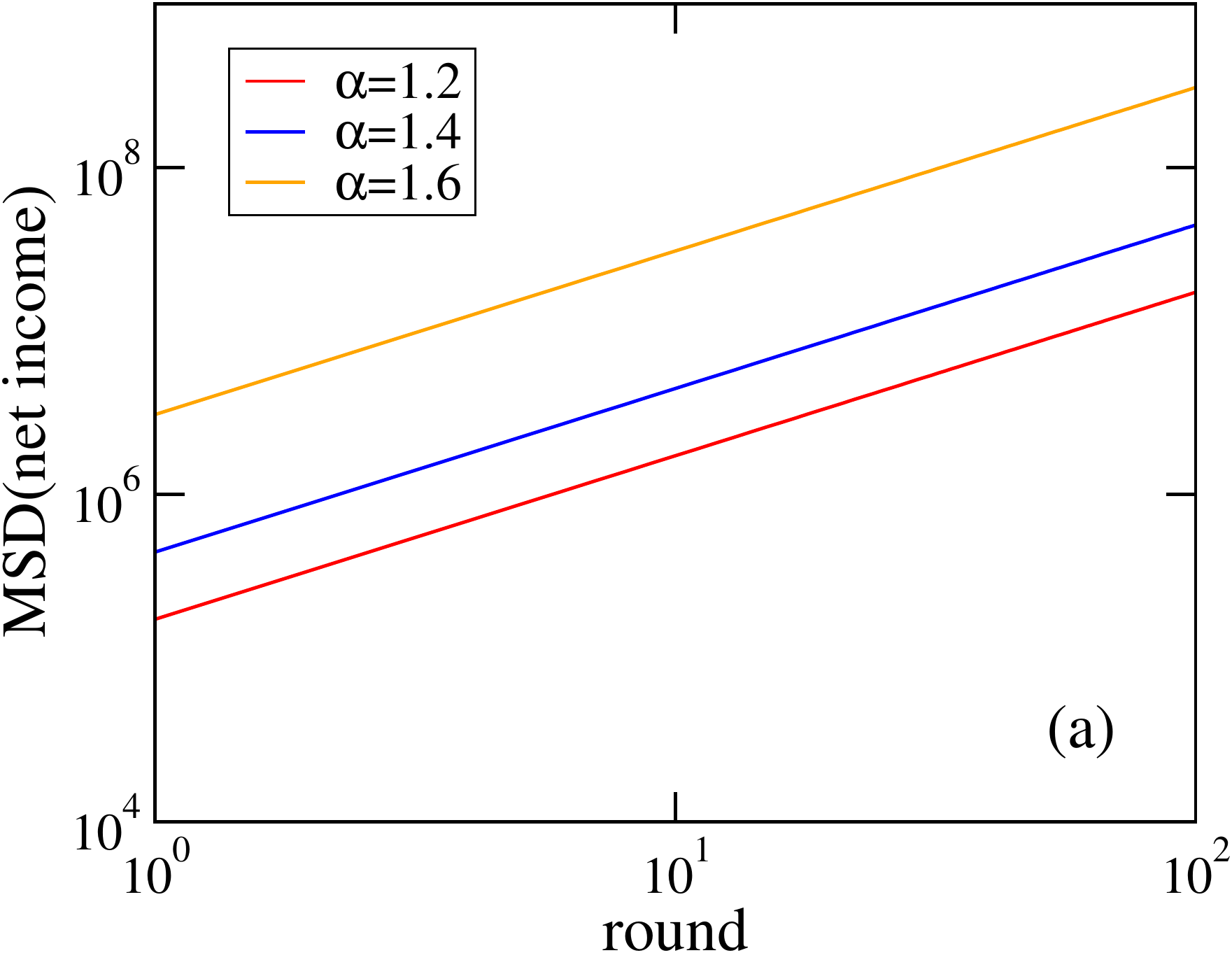}\quad
\includegraphics[width=0.4\columnwidth,clip=true]{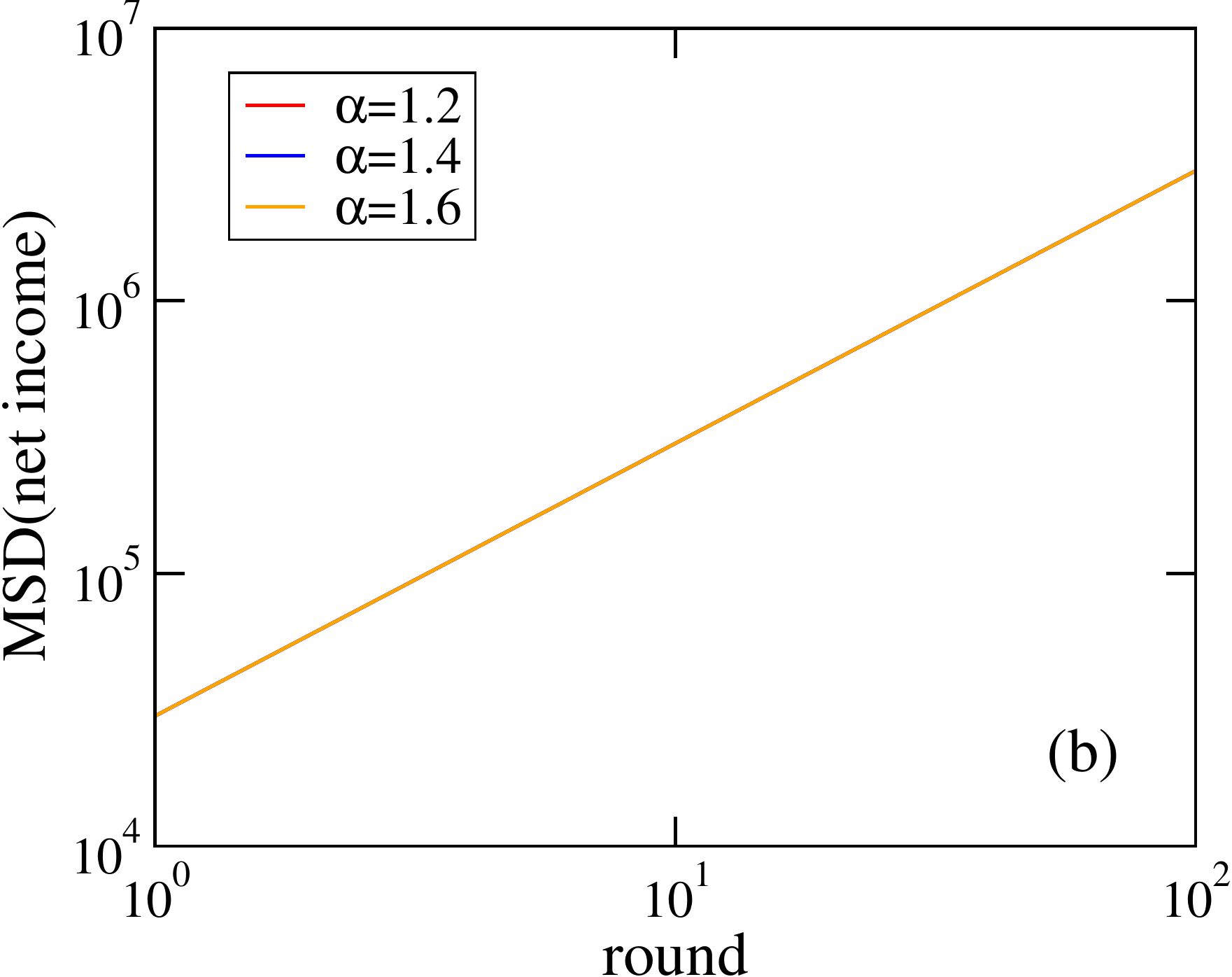}\\[0.5cm]
\includegraphics[width=0.4\columnwidth,clip=true]{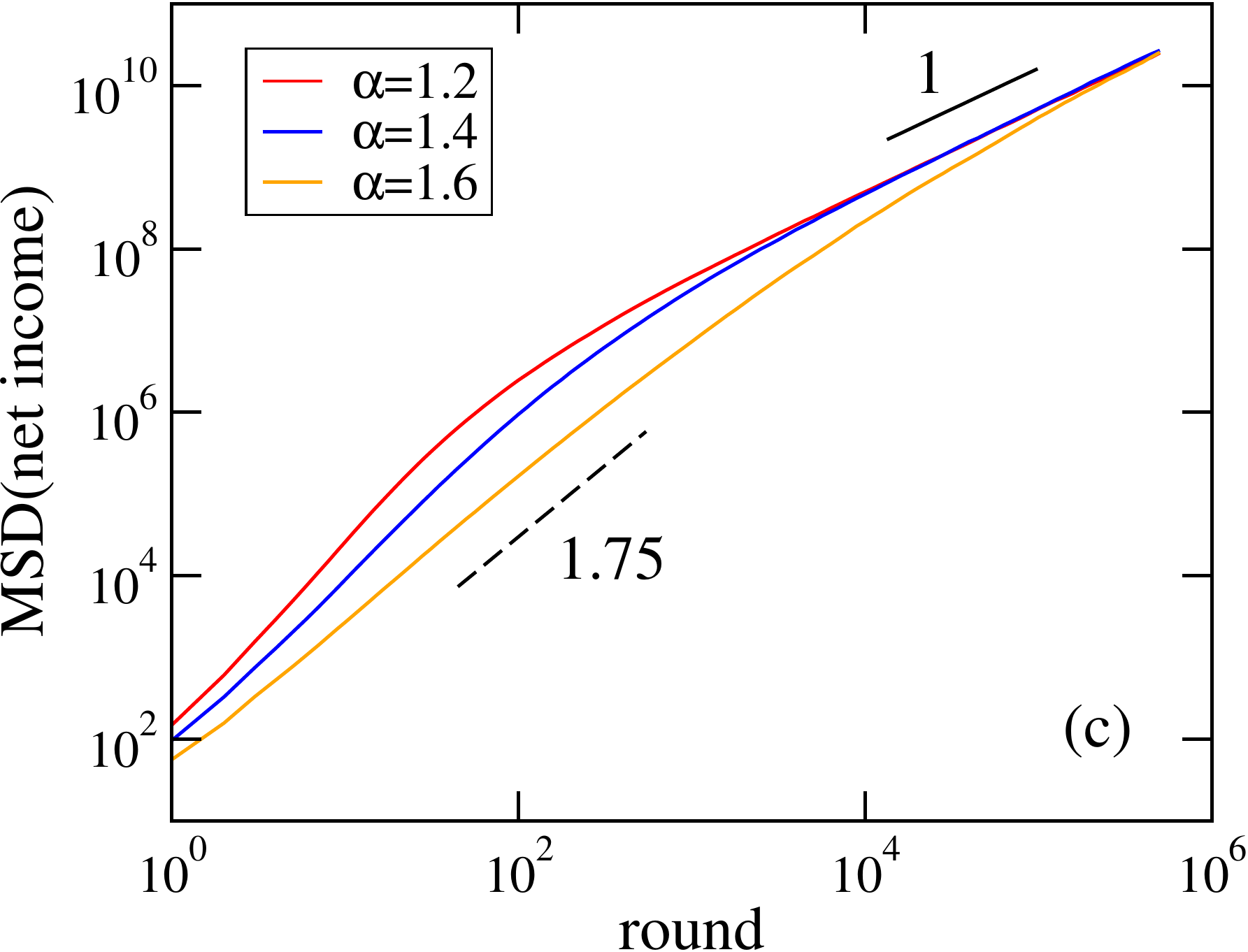}
\caption{The mean squared displacement for (a) model 1, (b) model 2, and (c) model 3. For models (1) and (2), the net income
of each gambler performs an independent random walk where the step length is related to the bet distribution (\ref{cont_power_exp}). In these two cases
the mean squared displacement increases linearly with time (i.e. the number of rounds played), in agreement with prior results.
Model 3, on the other hand, reveals a crossover from super-diffusion to diffusion. The different curves are for different values of the 
parameter $\alpha$ in the continuous power law distribution with an exponential cutoff (\ref{cont_power_exp}).
}
\label{fig13}
\end{figure}
%####################### Figure 13 #############################%

Similar to our analysis of the empirical data, we compute in the following for the different models 
the mean squared displacement (MSD) of the net income as well as the distribution of the first-passage time at which the income of
a gambler takes on a given target value. 

We start by noting that for models 1 and 2 the mean-squared displacement as a function of time
(i.e. the number of rounds played) can be computed exactly, see Appendix. For rounds
involving each time $n$ gamblers and a fixed mean bet value $\left< b \right>$,
the MSD is given for model 1 by
\begin{equation} \label{msd1}
\text{MSD}(t)=\left( \frac{2(n-1)}{n} \mu_2 + \frac{(n-1)(n-2)}{n} \left< b \right>^2 \right) t~,
\end{equation}
with $\mu_2$ being the second moment of the bet distribution,
whereas for model 2 one obtains
\begin{equation} \label{msd2}
\text{MSD}(t)=(n-1) \left< b \right>^2 t~.
\end{equation}
Fig. \ref{fig13}a and \ref{fig13}b display these curves for three different values of the parameter $\alpha$
found in the bet distribution (\ref{cont_power_exp}), with $\left< b \right>=100$ and $n=4$. 

Several comments are in order. First we note that although we consider a truncated power-law distribution, we obtain
that the MSD increases linearly with time. This is in agreement with an early observation of a linearly
increasing MSD encountered in simulations of truncated L\'{e}vy flights in two dimensions \cite{Ghaemi09}.
This linear time dependence is very general as the bet (i.e. step length) distribution only enters
through the mean and the second moment. Especially for model 2 any distribution with the same mean yields the same
MSD as expression (\ref{msd2}) does not depend on the variance. While we are focusing on the two truncated power-law
distributions (\ref{cont_power_exp}) and (\ref{cont_power}), even a distribution with finite mean and infinite second moment
yields for model 2 a finite MSD growing linearly with time. This is different for model 1 as the second moment
explicitly enters in expression (\ref{msd1}). As a result of this dependence, the MSDs for different values of $\alpha$,
see Fig. \ref{fig13}a, are shifted vertically, due to the fact that changing $\alpha$ while keeping $\left< b \right>$
constant changes the value of the second moment, see Appendix. We further note that these two models do not allow to obtain
a behavior similar to that observed in Fig. \ref{fig11} for the empirical data, namely a transition from a
super-diffusive behavior with an exponent larger than 1 to a normal diffusive behavior characterized by a linear increase
of the MSD. This, however, is different for model 3 where we indeed observe a crossover from super-diffusion
to normal diffusion, see Fig. \ref{fig13}c for data obtained for one million gamblers playing 50 million rounds,
with each round involving four randomly selected gamblers. As we can not compute the MSD analytically for this model, we can only provide
a heuristic argument for this observation. We note that in model 3 all gamblers have a finite wealth taken
from the distribution (\ref{wealth}). One of the consequences of this is that we add an absorbing boundary
(a gambler is removed once their wealth becomes zero),
another one is that initially many players have a small wealth and therefore can only bet small amounts. Consequently,
at early rounds the mean bet value of active players is smaller than the mean of the bet distribution (\ref{cont_power_exp}),
which makes the MSD to be smaller than what one obtains for models 1 and 2. As time increases, some gamblers are eliminated
as their wealth hits the absorbing boundary. As a result, the wealth of the active gamblers increases until their
mean bet values are getting close to the mean value of the bet distribution (\ref{cont_power_exp}). At this point
the MSD for model 3 shows a crossover from a super-diffusive behavior to a normal diffusive one.

%####################### Figure 14 #############################%
\begin{figure}
\includegraphics[width=0.4\columnwidth,clip=true]{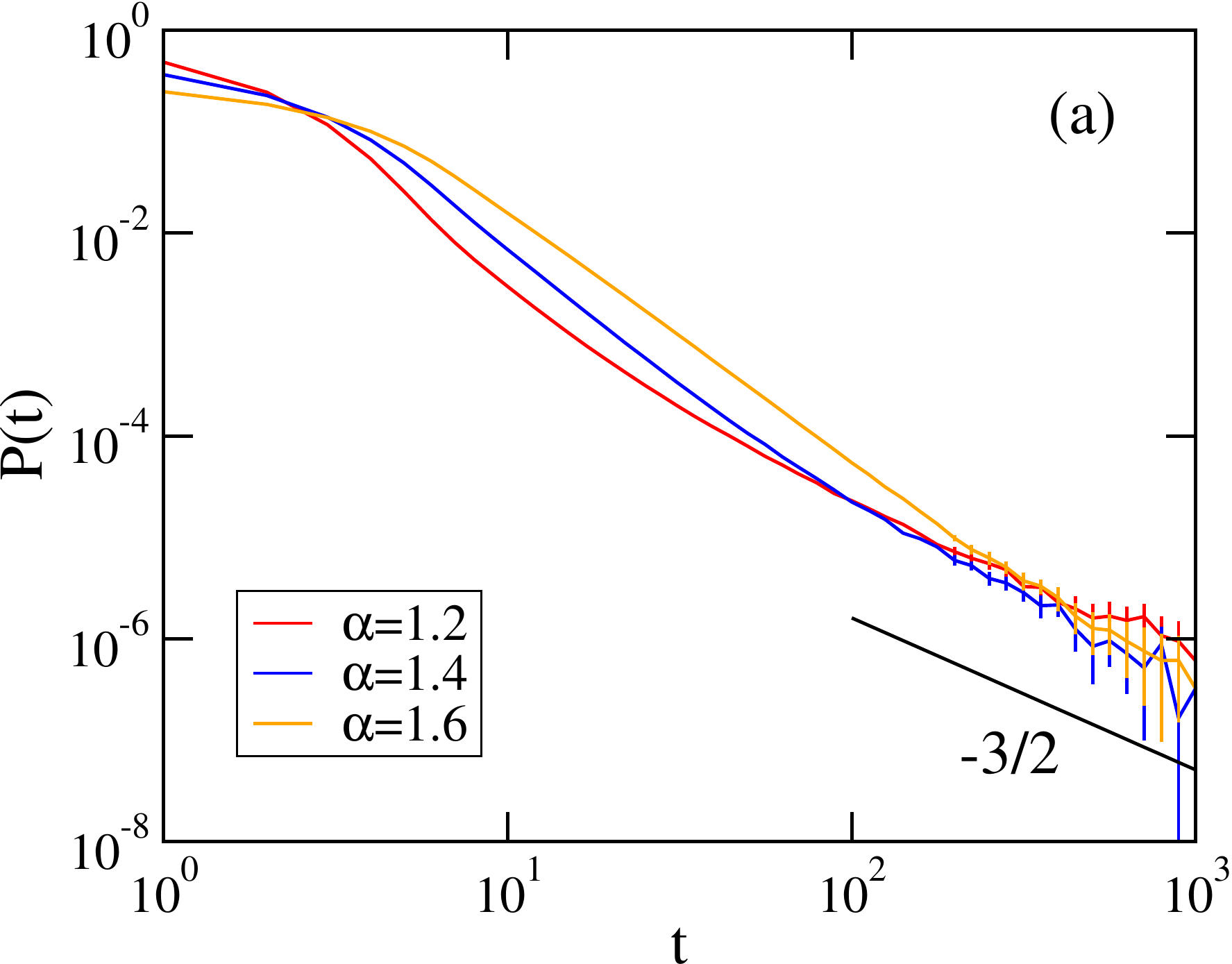}\quad
\includegraphics[width=0.4\columnwidth,clip=true]{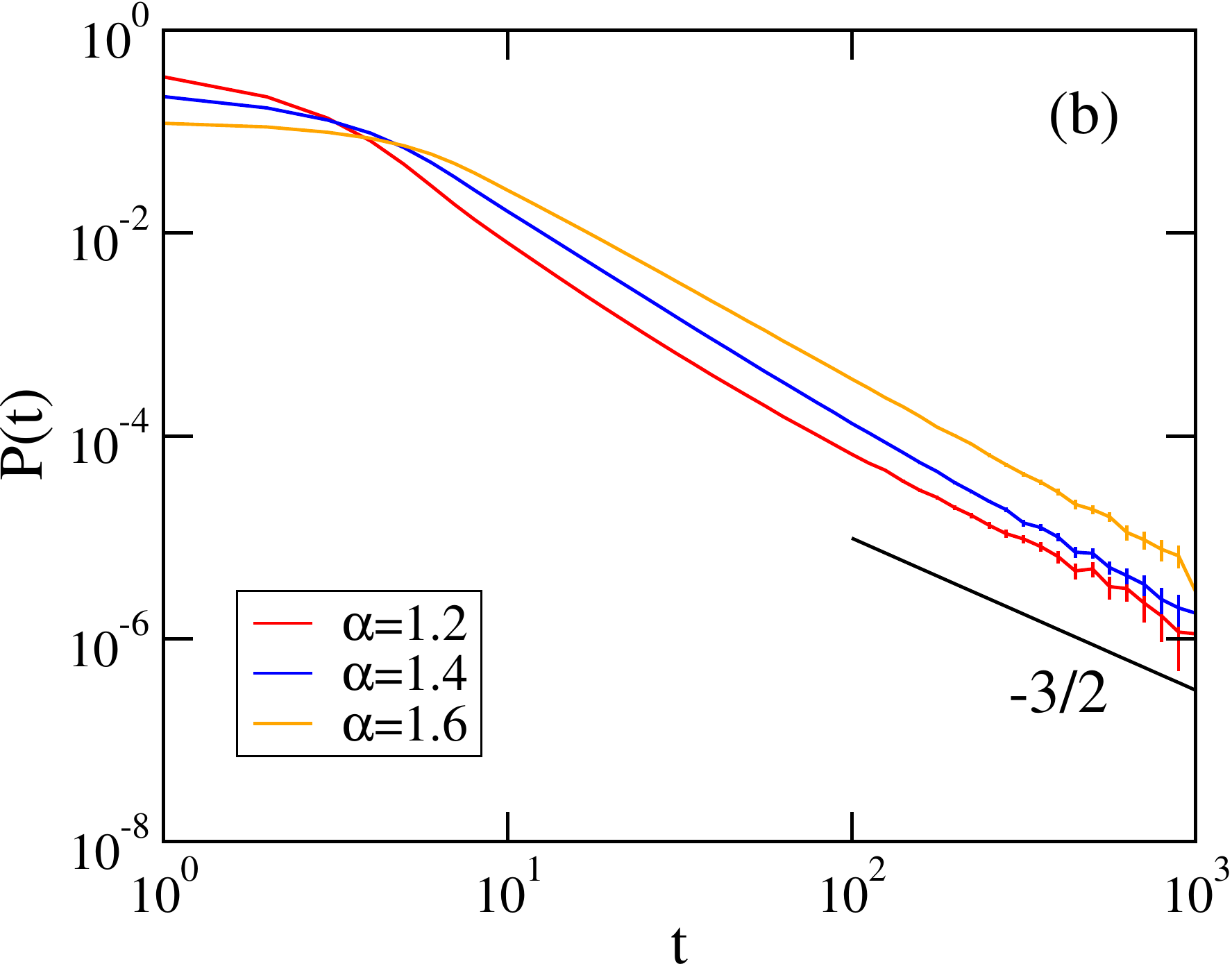}\\[0.5cm]
\includegraphics[width=0.4\columnwidth,clip=true]{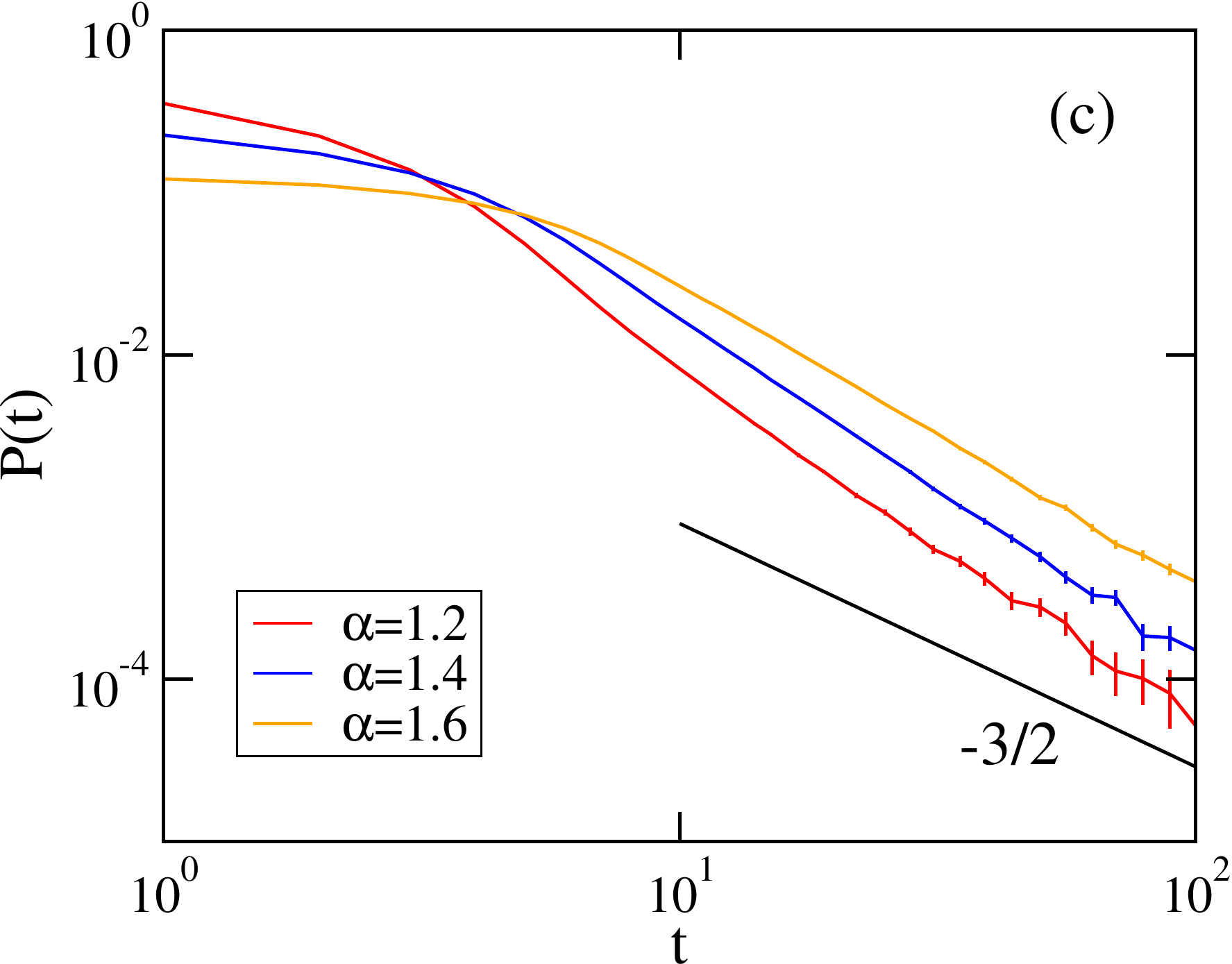}
\caption{The first-passage time distribution for (a) model 1, (b) model 2, and (c) model 3. For all three models we observe
a crossover from a super-diffusive behavior, revealed by a decay faster than $t^{-3/2}$, to a normal diffusive behavior 
proportional to $t^{-3/2}$.
Error bars indicate 95\% confidence intervals.
The different curves are for different values of the
parameter $\alpha$ in the bet distribution (\ref{cont_power_exp}).
}
\label{fig14}
\end{figure}
%####################### Figure 14 #############################%

Fig. \ref{fig14} shows our results for the first-passage time distributions obtained from simulations of the
three different models with the bet distribution (\ref{cont_power_exp}). 
For models 1 and 2 we simulate a gambler who plays 50 million rounds, which yields
a time series of their net income of length 50 million. The data shown in Fig. \ref{fig14}a and \ref{fig14}b 
result from averaging over 300 independent runs for $\alpha =1.2$, 250 independent runs for $\alpha =1.4$, and
75 independent runs for $\alpha =1.6$. These difference in the number of independent runs reflects an increase 
of computational costs when $\alpha$ increases, due to a decrease of the acceptance rates for generating
random numbers. For model 3 we only made one run with 1 million players and 50 million rounds. For
all three models we use $n=4$, $\left< b \right> = 100$, and $N_{fp}=20$.

Interestingly, all three models show in the first-passage time distribution the expected crossover from
a super-diffusive behavior at early times, characterized by a decay with an effective exponent larger than 3/2,
to a normal diffusive long-time behavior, where the distribution decays as $t^{-3/2}$. This crossover is rather
sharp for model 1, whereas it is more gradual for the other two models. As already mentioned in 
\cite{Inoue07}, the first-passage time distribution does not suffer from the same restrictions than
the MSD and is therefore the superior quantity for identifying the crossover between a super-diffusive
and a normal diffusive regime.

\section{Summary}
The quickly increasing video gaming industry has led to the development of other types of online entertainment, the
prime example being online gambling. We considered in this work an online jackpot game as an example of virtual item gambling.
Publicly available gambling logs permit a behavioral analysis at both the aggregate and individual levels. We analyzed
the probability distribution functions and correlation coefficients in order to elucidate
the relationships between some quantities derived from the gambling logs. Viewing the changes of the net income of a gambler as a random walk,
the mean squared displacement of the net income displays a transition from a super-diffusive to a diffusive behavior.
We discussed three different models, two of which are simple random walk models for a gambler with infinite wealth,
whereas the third one considers many gamblers with finite wealth. All three models show a crossover from super-diffusive
to normal diffusive behavior in the first-passage time distribution, but only the model with finite wealth displays
a similar crossover in the mean squared displacement.

\begin{acknowledgments}
This work is supported by the US National
Science Foundation through grant DMR-1606814.
\end{acknowledgments}

\appendix
\section{Mean squared displacements}
In the following we briefly discuss the exact calculation of the mean squared displacements for models 1 and 2.
In both models we consider $n$ players with infinite wealth who gamble with identically and independently distributed bet values $b$
taken from a distribution $P(b)$. In the main text we consider a power-law distribution (\ref{cont_power_exp}) as well as a 
power-law distribution with a sharp truncation (\ref{cont_power}).

Let $A, B, C, \cdots$ be the $n$ players attending one round. We are going to focus on player $A$ and compute the mean squared displacement of
their net income. For simplicity, we will also use $A, B, C, \cdots$ to represent the bet values of the corresponding players.
We denote by $A_1, A_2, \cdots, A_t$  the bet values of player $A$ in $t$ rounds and call $\Omega_1, \Omega_2, \cdots, \Omega_t$ the sum of the 
bet values of the other players in the corresponding rounds. We use $a_t$ to represent the net income of player $A$ after $t$ rounds and note that
before the first round played the net income is zero, i.e. $a_0 = 0$. The mean squared displacement is then given by
\begin{equation}
\text{MSD}(t)= \left< \left( a_t - a_0 \right)^2 \right> = \left<  a_t^2 \right>~.
\end{equation}

When player $A$ wins round $t$, then their net income increases by $\Omega_t = B_t + C_t + \cdots$, but the net income decreases by $- A_t$ in
case of a loss. Models 1 and 2 now differ by the probability to win the round, with this probability being given by $1/n$ for model 1 and by
$A_t/(B_t + C_t + \cdots)$ for model 2.

Let us first look at model 2. In that case the mean squared displacement at round $t$ is given by
\begin{equation*}
\begin{aligned}
\displaystyle \text{MSD}(t) = \int\limits_{A_1, \Omega_1, \cdots, A_t, \Omega_t} P\left(A_1, \Omega_1, \cdots, A_t, \Omega_t\right) \left( \frac{A_1}{A_1+\Omega_1}\cdots\frac{A_t}{A_t + \Omega_t}\left(\Omega_1 + \cdots+\Omega_t\right)^2\right.\\
+\frac{\Omega_1}{A_1+\Omega_1}\cdots\frac{A_t}{A_t + \Omega_t}\left(-A_1 + \cdots+\Omega_t\right)^2 + \cdots + \frac{A_1}{A_1+\Omega_1}\cdots\frac{\Omega_t}{A_t + \Omega_t}\left(\Omega_1 + \cdots-A_t\right)^2\\
\left. + \frac{\Omega_1}{A_1+\Omega_1}\cdots\frac{\Omega_t}{A_t + \Omega_t}\left(-A_1 - \cdots-A_t\right)^2 \right) \rm{d} A_1 \rm{d} \Omega_1 \cdots \rm{d} A_t \rm{d} \Omega_t.
\end{aligned}
\end{equation*}
After expanding the squared terms most terms cancel out, yielding after some simple algebraic manipulations
\begin{eqnarray}
\displaystyle \text{MSD}(t) & = & \int\limits_{A_1, \Omega_1, \cdots, A_t, \Omega_t} P\left(A_1, \Omega_1, \cdots, A_t, \Omega_t\right) \left(A_1\Omega_1 + \cdots + A_t\Omega_t \right) \rm{d} A_1 \rm{d} \Omega_1 \cdots \rm{d} A_t \rm{d} \Omega_t \nonumber \\
& = & \int\limits_{A_1, \Omega_1} P(A_1, \Omega_1) A_1 \Omega_1 \rm{d} A_1 \rm{d} \Omega_1 + \cdots + \int\limits_{A_t, \Omega_t} P(A_t, \Omega_t) A_t \Omega_t \rm{d} A_t \rm{d} \Omega_t \nonumber
\end{eqnarray}
All these $t$ terms are identical and are given by
\begin{eqnarray}
\displaystyle \int\limits_{A, \Omega} P(A, \Omega) A \Omega \rm{d} A \rm{d} \Omega
& = & \int\limits_{A, B, C, \cdots} P(A) P(B) P(C) \cdots \left( A (B + C + \cdots)\right) \rm{d} A \rm{d} B \rm{d} C \cdots \nonumber \\
& = & \int\limits_{A} P(A) A \rm{d} A \left( \int\limits_{B} P(B) B \rm{d} B + \int\limits_{C} P(C) C \rm{d} C + \cdots \right) \nonumber \\
& = & \mu \left( (n-1) \mu \right) = (n-1) \mu^2 \nonumber
\end{eqnarray}
where $\mu = \left< A \right> = \int\limits_{A} P(A) A \rm{d} A$. This then yields the final result
$$\text{MSD}(t) = (n-1) \left< A \right>^2 + \cdots + (n-1) \left< A \right>^2 = (n-1) \left< A \right>^2 t.$$
We note that for model 2 the MSD grows linearly in time. As the MSD is independent of the second moment
of the bet distribution, it is the same for any bet distribution with the same mean, including distributions with 
finite mean and infinite second moment.

The calculation for model 1 closely follows that of model 2, but with the major change that for gambler A the probability of winning a round is $1/n$,
whereas the probability of losing that round is $(n-1)/n$, with $n$ being the number of gamblers involved in a round. This then yields the expression
\begin{equation*}
\begin{aligned}
\displaystyle \text{MSD}(t) = \langle a_t^2 \rangle = \int\limits_{A_1, \Omega_1, \cdots, A_t, \Omega_t} P\left(A_1, \Omega_1, \cdots, A_t, \Omega_t\right) \left( \frac{1}{n}\cdots\frac{1}{n}\left(\Omega_1 + \cdots+\Omega_t\right)^2\right.\\
+\frac{n-1}{n}\cdots\frac{1}{n}\left(-A_1 + \cdots+\Omega_t\right)^2 + \cdots + \frac{1}{n}\cdots\frac{n-1}{n}\left(\Omega_1 + \cdots-A_t\right)^2\\
\left. + \frac{n-1}{n}\cdots\frac{n-1}{n}\left(-A_1 + \cdots-A_t\right)^2 \right) \rm{d} A_1 \rm{d} \Omega_1 \cdots \rm{d} A_t \rm{d} \Omega_t~,
\end{aligned}
\end{equation*}
and, after some algebraic manipulations,
\begin{eqnarray}
\displaystyle \text{MSD}(t) & = & \int\limits_{A_1, \Omega_1, \cdots, A_t, \Omega_t} P\left(A_1, \Omega_1, \cdots, A_t, \Omega_t\right) \left(\frac{n-1}{n} A_1^2 + \frac{1}{n} \Omega_1^2 + \cdots + \frac{n-1}{n} A_t^2 + \frac{1}{n}\Omega_t^2 \right) \nonumber \\
&& \hspace*{11cm} \rm{d} A_1 \rm{d} \Omega_1 \cdots \rm{d} A_t \rm{d} \Omega_t \nonumber \\
& = & \int\limits_{A_1, \Omega_1} P(A_1, \Omega_1) \left(\frac{n-1}{n} A_1^2 + \frac{1}{n} \Omega_1^2\right) \rm{d} A_1 \rm{d} \Omega_1 + \cdots 
 \int\limits_{A_t, \Omega_t} P(A_t, \Omega_t) \left(\frac{n-1}{n} A_t^2 + \frac{1}{n} \Omega_t^2\right) \rm{d} A_t \rm{d} \Omega_t \nonumber 
\end{eqnarray}
Again, these $t$ terms are identical, with
\begin{eqnarray}
&& \int\limits_{A, \Omega} P(A, \Omega) \left(\frac{n-1}{n} A^2 + \frac{1}{n} \Omega^2\right) \rm{d} A \rm{d} \Omega \nonumber \\
& = & \int\limits_{A, B, C, \cdots} P(A, B, C, \cdots) \left(\frac{1}{n} (B + C + \cdots)^2 + \frac{n-1}{n} A^2 \right) \rm{d} A \rm{d} B \rm{d} C \cdots \nonumber \\
& = & \frac{n-1}{n}\int\limits_{A} P(A) A^2 \rm{d} A + \frac{1}{n} \int\limits_{B+C+\cdots} P(B) P(C) \cdots \left( B^2 + C^2 + 2BC + \cdots\right) \rm{d} B \rm{d} C \cdots \nonumber \\
& = & \frac{n-1}{n}\mu_2 + \frac{1}{n}\left( \int\limits_{B} P(B) B^2 \rm{d} B + \int\limits_{C} P(C) C^2 \rm{d} C+ \cdots\right) +\frac{1}{n} \left(\int\limits_{B, C} P(B) P(C)\ 2 B C\ \rm{d} B \rm{d} C + \cdots\right) \nonumber \\
& = & \frac{n-1}{n}\mu_2 + \frac{1}{n} (n-1) \mu_2 + 
\frac{1}{n}\frac{(n-1)(n-2)}{2} 2\mu^2 \nonumber \\
& = & \frac{2(n-1)}{n}\mu_2 + \frac{(n-1)(n-2)}{n} \mu^2 \nonumber
\end{eqnarray}
with the mean $\mu = \left< A \right> = \int\limits_{A} P(A) A \rm{d} A$ and the second moment $\mu_2 = \left< A^2 \right>
= \int\limits_{A} P(A) A^2 \rm{d} A$.
It follows that the MSD for model 1 is given by
\begin{displaymath}
\text{MSD}(t) = \left(\frac{2(N-1)}{N}\mu_2 + \frac{(N-1)(N-2)}{N} \mu^2 \right) t
\end{displaymath}
and is therefore still proportional to $t$, but now the pre-factor depends on both the mean and the second moment.

The Table below provides the interested reader with the mean values and second moments for the two bet distributions considered in this work.
$\Gamma \left( \cdot,\cdot \right)$
is the incomplete Gamma function. 

\begin{table}[ht]
\centering
\label{table2}
\begin{tabular}{|c|c|c|}
\hline
distribution model                & mean $\mu$                                                                                                                    & second moment $\mu_2$                                                                                                           \\ \hline
power law with exponential cutoff (\ref{cont_power_exp})& $\displaystyle \frac{1}{\lambda} \frac{\Gamma(2-\alpha, \lambda b_{min})}{\Gamma(1-\alpha, \lambda b_{min})}$                 & $\displaystyle \frac{1}{\lambda^2} \frac{\Gamma(3-\alpha, \lambda b_{min})}{\Gamma(1-\alpha, \lambda b_{min})}$               \\ \hline
power law with sharp truncation (\ref{cont_power})  & $\displaystyle \frac{1-\alpha}{2-\alpha} \frac{b_{min}^{2-\alpha}-b_{max}^{2-\alpha}}{b_{min}^{1-\alpha}-b_{max}^{1-\alpha}}$ & $\displaystyle \frac{1-\alpha}{3-\alpha} \frac{b_{min}^{3-\alpha}-b_{max}^{3-\alpha}}{b_{min}^{1-\alpha}-b_{max}^{1-\alpha}}$ \\ \hline
\end{tabular}
\caption{Mean and second moment for the different bet value distributions.}
\end{table}

\end{document}